\documentclass{aa}  

\usepackage{graphicx}
\usepackage{txfonts}
\usepackage{xcolor}
\usepackage{multicol}
\def\PLUTO{{\sc pluto}}

\newcommand\rs[1]{_\mathrm{#1}}

\newcommand{\casa}{Cas~A}

\begin{document} 


    
\title{Origin of holes and rings in the Green Monster of Cassiopeia~A: Insights from 3D magnetohydrodynamic simulations}


   \author{S.\,Orlando\inst{1}
      \and H.-T.\,Janka\inst{2}
      \and A.\,Wongwathanarat\inst{2}
      \and F.\,Bocchino\inst{1}
      \and I.\,De Looze\inst{3}
      \and D.\,Milisavljevic\inst{4}
      \and M.\,Miceli\inst{5,1}
      \and \\ T.\,Temim\inst{6}
      \and J.\,Rho\inst{7,8}
      \and S.\,Nagataki\inst{9,10,11}
      \and M.\,Ono\inst{12,9}
      \and V.\,Sapienza\inst{1}
      \and E.\,Greco\inst{1}
  }

\institute{INAF -- Osservatorio Astronomico di Palermo, Piazza del Parlamento 1, I-90134 Palermo, Italy\\
\email{salvatore.orlando@inaf.it}
\and Max-Planck-Institut f\"ur Astrophysik, Karl-Schwarzschild-Str. 1, D-85748 Garching, Germany
\and Sterrenkundig Observatorium, Ghent University, Krijgslaan 281 - S9, B-9000 Gent, Belgium
\and Department of Physics and Astronomy, Purdue University, 525 Northwestern Avenue, West Lafayette, IN 47907, USA
\and Dip. di Fisica e Chimica, Universit\`a degli Studi di Palermo, Piazza del Parlamento 1, 90134 Palermo, Italy
\and Princeton University, 4 Ivy Ln, Princeton, NJ 08544, USA
\and SETI Institute, 339 Bernardo Avenue, Suite 200, Mountain View, CA 94043, USA
\and Department of Physics and Astronomy, Seoul National University, Gwanak-ro 1, Gwanak-gu, Seoul, 08826, Republic of Korea
\and Astrophysical Big Bang Laboratory (ABBL), RIKEN Cluster for Pioneering Research, 2-1 Hirosawa, Wako, Saitama 351-0198, Japan
\and RIKEN Interdisciplinary Theoretical \& Mathematical Science Program (iTHEMS), 2-1 Hirosawa, Wako, Saitama 351-0198, Japan
\and Astrophysical Big Bang Group (ABBG), Okinawa Institute of Science and Technology (OIST), 1919-1 Tancha, Onna-son, Kunigami-gun, Okinawa 904-0495, Japan
\and Institute of Astronomy and Astrophysics, Academia Sinica, No.1, Sec.4, Roosevelt Road, Taipei 106319, Taiwan
             }

   \date{Received XXX; accepted XXX}

 
  \abstract
   {The supernova remnant (SNR) Cassiopeia A (\casa) offers a unique opportunity to study supernova (SN) explosion dynamics and remnant interactions with the circumstellar medium (CSM). Recent observations with the James Webb Space Telescope (JWST) have unveiled an enigmatic structure within the remnant, termed "Green Monster" (GM), whose properties indicate a CSM origin.}
   {Our goal is to investigate the properties of the GM and uncover the origin of its intriguing pockmarked structure, characterized by nearly circular holes and rings. We aim to examine the role of small-scale ejecta structures in shaping these features through their interaction with a dense circumstellar shell.}
   {We adopted a neutrino-driven SN model to trace the evolution of its explosion from core collapse to the age of the \casa\ remnant using high-resolution 3D magnetohydrodynamic simulations. Besides other processes, the simulations include self-consistent calculations of radiative losses, accounting for deviations from electron-proton temperature equilibration and ionization equilibrium, as well as the ejecta composition derived from the SN.}
   {The observed GM morphology can be reproduced by the interaction of dense ejecta clumps and fingers with an asymmetric, forward-shocked circumstellar shell. The clumps and fingers form by hydrodynamic instabilities growing at the interface between SN ejecta and shocked CSM. Radiative cooling accounting for effects of non-equilibrium of ionization enhances the ejecta fragmentation, forming dense knots and thin filamentary structures that penetrate the shell, producing a network of holes and rings with properties similar to those observed.}
   {The origin of the holes and rings in the GM can be attributed to the interaction of ejecta with a shocked circumstellar shell. By constraining the timing of this interaction and analyzing the properties of these structures, we provide a distinction of this scenario from an alternative hypothesis, which attributes these features to fast-moving ejecta knots penetrating the shell ahead of the forward shock.}


\titlerunning{Origin of holes and rings in the Green Monster of Cassiopeia~A}
\authorrunning{S. Orlando et~al.}

   \maketitle
%

\section{Introduction}

The remnants of core-collapse supernovae (SNe) appear as diffuse, extended sources with highly intricate morphologies, shaped by the interplay of shocks and the complex spatial distribution of ejecta originating from the progenitor star. In young supernova remnants (SNRs; less than 1000 years old), these features provide a unique window into several key aspects of the explosion mechanism. They reveal nucleosynthetic yields, highlight large-scale asymmetries arising from the earliest phases of the explosion, and offer insights into the processes at work in the aftermath of core collapse (e.g., \citealt{2015A&A...577A..48W, 2017ApJ...842...13W, 2016ApJ...822...22O, 2021A&A...645A..66O}). Furthermore, they can shed light on the nature of the progenitor star and the structure of the surrounding circumstellar medium (CSM), which reflects the star's mass-loss history (e.g., \citealt{2021A&A...654A.167U, 2020A&A...636A..22O, 2024ApJ...977..118O}). As a result, SNRs encode invaluable information about the final stages of massive star evolution, offering critical clues about the poorly understood mechanisms governing mass loss in progenitor stars during the approach to core collapse (e.g., \citealt{2014ARA&A..52..487S} and references therein).

Cassiopeia A (\casa), a prototypical core-collapse SNR located approximately 3.4 kpc from Earth (\citealt{1995ApJ...440..706R}), has been extensively studied across the electromagnetic spectrum, serving as a cornerstone for understanding the physics of core-collapse SNe. Its young age, roughly 350 years (e.g., \citealt{2001AJ....122..297T, 2006ApJ...645..283F}), combined with its relatively close proximity, makes it a unique laboratory for probing the expansion of the SN explosion and its interaction with the CSM. 

Indeed, recent observations with the James Webb Space Telescope (JWST) have provided groundbreaking insights into \casa\ (\citealt{2024ApJ...965L..27M}). The analysis of carbon monoxide (CO) formation and destruction has unveiled intricate spatial and velocity structures of CO emission in the ejecta, highlighting its re-formation behind the reverse shock and offering valuable insights into the interplay between molecules, dust, and highly ionized ejecta (\citealt{2024ApJ...969L...9R}). For the first time, a web-like network of unshocked ejecta filaments has been detected (\citealt{2024ApJ...965L..27M}, Dickinson et al., in preparation).  These filaments likely retain a "memory" of the early explosion conditions, tracing key processes such as neutrino-heated bubble expansion immediately after core collapse, hydrodynamic (HD) instabilities during blast propagation through the stellar interior, and the Ni-bubble effect following shock breakout, mechanisms that collectively shaped the ejecta structure during and shortly after the SN event (\citealt{2025arXiv250300130O}). Furthermore, a striking structure in the near-side central region of the remnant, known as the "Green Monster" (GM), has been revealed (\citealt{2024ApJ...965L..27M, 2024ApJ...976L...4D, 2024ApJ...964L..11V}), adding further complexity to our understanding of the early evolution and morphology of \casa.

The GM is a prominent emission feature identified in the mid-infrared (IR) band by composite MIRI F1130W and F1280W images. This structure is dominated by dust and emission lines of neon (Ne), hydrogen (H), and iron (Fe) at low radial velocities, leading to its association with CSM likely ejected during an asymmetric mass-loss phase of the progenitor star (\citealt{2024ApJ...976L...4D}). Interestingly, \cite{2022A&A...666A...2O} anticipated the \casa's interaction with a dense, asymmetric circumstellar shell, occurring roughly within the last 100 years, to explain some of the enigmatic asymmetries observed, for instance, in the remnant's reverse shock evolution (see also \citealt{2022ApJ...929...57V, 2025arXiv250107708F}). They further suggested that this shell might have resulted from a massive eruption of the progenitor star between $10^4$ and $10^5$ years prior to the SN explosion. Decoding the origin and properties of the GM may therefore provide crucial insights into the mass-loss history of the \casa\ progenitor and its final evolutionary stages. 

A puzzling characteristic of the GM is its array of circular holes, measuring $1''-3''$ in size and encircled by bright rings (\citealt{2024ApJ...976L...4D}). The precise origin of these structures remains uncertain, with various hypotheses explored. These include: (i) the expansion of small Ni-rich ejecta embedded in the GM, (ii) interactions of high-velocity ejecta filaments with dense circumstellar clumps, (iii) impacts of small, high-velocity ejecta "bullets" with the GM prior to forward-shock contact, and (iv) interaction of the GM with ejecta clumps and fingers after the forward shock's passage.

Among the proposed scenarios, the last two are the most promising, offering a plausible explanation for the formation of the holes and rings (\citealt{2024ApJ...976L...4D}). In particular, the interaction of the GM with ejecta after the forward shock's passage was suggested by 3D HD and magneto-hydrodynamic (MHD) simulations modeling the interaction of \casa\ with an asymmetric circumstellar shell (\citealt{2022A&A...666A...2O}). In these simulations, holes in the shell form as a result of interactions between the dense, shocked shell and ejecta filaments driven by HD instabilities (i.e., Rayleigh-Taylor, Richtmyer-Meshkov, and Kelvin-Helmholtz shear instabilities; \citealt{1973MNRAS.161...47G, 1991ApJ...367..619F, 1992ApJ...392..118C, 1996ApJ...472..245J}). These instabilities generate filamentary structures of shocked ejecta or "fingers", which extend from the contact discontinuity toward the forward shock. As these ejecta fingers protrude into the dense shell material, they produce the observed holes.

However, a main difference between the simulated and observed holes lies in their size, with the former having a radius roughly three times larger than the latter (see Fig.~10 in \citealt{2024ApJ...976L...4D}). This discrepancy may be due to limitations in the spatial resolution of the simulations, which could be insufficient to capture the small-scale structures of the ejecta. Additionally, the omission of certain physical processes, such as radiative losses from optically thin plasma, in previous simulations of \casa\ may contribute to the differences between the modeled and observed structures. 

To address this issue and contribute to unraveling the origin of the GM and its pockmarked structure, we adopted the 3D MHD model from \cite{2022A&A...666A...2O}, which simulates the evolution of a neutrino-driven SN explosion from the immediate post-core-collapse phase to its current state. This model includes the interaction with a circumstellar shell, replicating key characteristics of \casa. Building on this framework, we conducted high-resolution simulations to capture the finer details of ejecta structures at small scales as revealed by JWST. In addition, the simulations incorporate radiative losses from optically thin plasma, accounting for deviations from both electron-ion temperature equilibration and ionization equilibrium within the emitting plasma in each computational cell. Radiative losses were calculated based on the plasma composition in each cell, including the ejecta abundances derived from the SN explosion. The primary goal of this study is to explore how small-scale ejecta structures and radiative cooling influence the observed features of the GM, as well as to determine the timing of the interaction between the ejecta and the circumstellar shell. These results are expected to provide new insights into the nature of the GM and offer valuable constraints on the mass-loss history of \casa's progenitor star.

The paper is organized as follows: Sect.~\ref{sec:model} provides an overview of the numerical setup employed for our 3D HD and MHD simulations; Sect.~\ref{sec:results} presents the simulation results and compares them with the observed properties of the GM in \casa; Sect.~\ref{sec:summary} summarizes our results and discusses their implications for the late evolutionary stages of the progenitor star of \casa. Finally, Appendix~\ref{app:losses} details the implementation of the numerical module used to compute radiative losses from optically thin plasma as a function of electron temperature, ionization age, and plasma abundances; in Appendix~\ref{app:multi-media}, we describe the content of online multi-media material, offering interactive and dynamic representations of the
phenomena discussed in the paper.

\section{The 3D MHD model}
\label{sec:model}

The numerical framework used in this study has been described in previous works (\citealt{2021A&A...645A..66O, 2022A&A...666A...2O}). It models the evolution of a neutrino-driven SN explosion, starting from the initial core collapse and extending through the formation and expansion of the remnant over approximately 1000 years. Specifically, we used the W15-2-cw-IIb neutrino-driven explosion model developed by \cite{2017ApJ...842...13W}, and selected for its ability to replicate several key properties observed in \casa. This model was used as initial condition for high-resolution HD and MHD simulations which describe the transition to the SNR phase (\citealt{2021A&A...645A..66O}). The simulations track the remnant's expansion and its interaction with the complex, inhomogeneous CSM, including the presence of a dense circumstellar shell (\citealt{2022A&A...666A...2O}). Below, we summarize the main components and physical processes incorporated into the model. For a detailed description of the implementation, we refer the reader to the aforementioned papers.

\subsection{Supernova explosion model}

The SN model adopted (W15-2-cw-IIb) was neither designed nor specifically tuned to replicate \casa. The progenitor star was initially a $15\,M_\odot$ star (model W15; \citealt{1995ApJS..101..181W}) whose neutrino-driven explosion (see \citealt{2015A&A...577A..48W}) stochastically developed a nickel (Ni) ejecta geometry resembling the shocked Fe morphology observed in \casa. While this geometry is not uncommon for neutrino-driven explosions, considerable variability exists depending on factors such as the progenitor properties, explosion energy, and stochastic effects. Unipolar, bipolar, and more spherical multi-polar large-scale structures are all possible outcomes (see \citealt{2015A&A...577A..48W} for more examples).

Model W15-2-cw-IIb has been selected because its explosion energy, helium (He) core mass (which determines the ejecta mass), and large-scale asymmetries formed shortly after core collapse closely align with key observed properties of \casa\ (see \citealt{2017ApJ...842...13W}). The progenitor was modeled as having been stripped of most of its H envelope to simulate a Type IIb SN, consistent with evidence linking \casa\ to this SN type \citep{2008Sci...320.1195K, 2011ApJ...732....3R}. Extensive pre-explosion mass loss was assumed, leaving the progenitor with a residual H envelope of approximately $0.3\,M_\odot$ \citep{2017ApJ...842...13W}. The explosion was driven by neutrino energy deposition, releasing $\sim 1.5 \times 10^{51}$ erg and ejecting $3.3\,M_\odot$ of stellar material into the surrounding medium (see Table~\ref{Tab:model}). While the ejecta mass is consistent with observational estimates, the explosion energy is slightly below values inferred from observations, which suggest $\sim 2 \times 10^{51}$ erg (e.g., \citealt{2003ApJ...597..347L, 2020ApJ...893...49S}).

The 3D SN simulation tracks the evolution from approximately 15 milliseconds after the core bounce to well beyond the breakout of the shock front at the stellar surface (occurring at $\sim$1500\,s after bounce) until roughly one day later. Key physical processes include self-gravity, neutron star formation, fallback material, and a Helmholtz equation of state for accurate thermodynamic properties. Additionally, the model includes an $\alpha$-network for nucleosynthesis, providing approximate insights into the chemical composition of the ejecta. Asymmetries in the ejecta naturally arise from stochastic processes such as convective overturn due to neutrino heating and the standing accretion shock instability (SASI) in the first second after bounce (e.g., \citealt{2003ApJ...584..971B, 2017hsn..book.1095J, 2021Natur.589...29B}).

\subsection{Transition to the SNR phase}

The output of the SN model at $\approx 17.85$~hours after core collapse serves as the initial condition for the 3D HD and MHD simulations, which track the transition to the remnant phase. The CSM is modeled as a stellar wind with a density profile proportional to $r^{-2}$, normalized to $0.8$~cm$^{-3}$ at a radius of 2.5 pc according to constraints from observations (\citealt{2014ApJ...789....7L}). To replicate the significant asymmetries observed in the reverse shock of \casa\ \citep{2022ApJ...929...57V, 2025arXiv250107708F}, the simulations incorporate an asymmetric, dense circumstellar shell, with its densest region positioned on the near side to the northwest (see \citealt{2022A&A...666A...2O} for more details on the description of the shell). The MHD simulations describe a Parker-spiral magnetic field, reflective of the progenitor's stellar wind dynamics. In particular, this field geometry reflects the combined effect of the progenitor’s rotation and mass loss (\citealt{1958ApJ...128..664P}; see \citealt{2019A&A...622A..73O} for further details). 

To accurately capture the complex physics of SNR evolution, several key processes are incorporated into the model: (i) the energy deposition from the radioactive decay chain $^{56}$Ni $\rightarrow$ $^{56}$Co $\rightarrow$ $^{56}$Fe, treated as a local energy source, excluding neutrino contributions and assuming no $\gamma$-ray leakage from the inner part of the remnant \citep{2021A&A...645A..66O}; (ii) the deviations from equilibrium of ionization, accounted for by calculating the maximum ionization parameter, $\tau$, in each cell of the computational domain \citep{2015ApJ...810..168O}; (iii) the deviations from electron-proton temperature equilibration, with post-shock temperatures calculated by Coulomb collisions (\citealt{2015ApJ...810..168O}).

In addition to processes included in previous simulations, the current model includes also radiative losses from optically thin plasma. These losses, $\Lambda (T_{\rm e}, \tau, Z)$, are calculated in each cell as a function of the electron temperature ($T_{\rm e}$), ionization age ($\tau$), and plasma abundances ($Z$). This self-consistent approach ensures that radiative losses reflect the actual physical and chemical conditions of the emitting plasma, which is especially important for the shocked ejecta. The ejecta composition is directly inherited from the nucleosynthesis outputs of the SN model. In Appendix \ref{app:losses}, we detail the numerical implementation of these radiative losses. We note that radiative losses become significant for plasma temperatures exceeding $10^4$~K, influencing the evolution of shocked ejecta and shocked shell. On the other hand, they have little effect on the inner unshocked ejecta, including the network of O-rich ejecta filaments revealed by JWST (\citealt{2024ApJ...965L..27M, 2025arXiv250300130O}, Dickinson et al., in preparation).

It is worth noting that in our simulations, we did not account for the additional radiative cooling caused by gas-grain collisions. In regions where gas and dust coexist, collisions between gas particles (such as electrons, ions, and neutral atoms) and dust grains transfer kinetic energy from the gas to the grains. This energy heats the dust grains, which then radiate it as thermal emission, predominantly in the IR part of the spectrum. These radiative losses from dust grains provide an additional cooling mechanism, complementing the radiative cooling process discussed above. This enhanced cooling can lead to increased compression and fragmentation of the shocked ejecta, potentially producing thinner fingers and smaller clumps of ejecta than modeled here. However, the overall impact of this process is expected to be limited. This cooling mechanism is most effective in the shocked ejecta, where dust grains are subject to destruction by the reverse shock, significantly reducing their contribution to radiative cooling.

The SNR simulations were performed using the \PLUTO\ code, a highly versatile numerical tool for astrophysical fluid dynamics (\citealt{2007ApJS..170..228M, 2012ApJS..198....7M}). For HD calculations, the Roe Riemann solver was adopted, while the HLLD solver was used for MHD simulations. The computational domain consists of a Cartesian grid with $(2048)^3$ cells, dynamically expanding to track the remnant's evolution from about one day after the breakout of the shock wave at the stellar surface to the SNR phase, up to 1000 years post-explosion. This approach allows the numerical grid to expand as the forward shock propagates outward, ensuring adequate resolution across all evolutionary stages. Consecutive remappings were performed when the forward shock approaches the boundary of the domain to optimize computational efficiency while maintaining accuracy (see \citealt{2021A&A...645A..66O} for details). At the current age of \casa\ (350 years), the spatial resolution reached 0.002 pc, namely twice as fine as in previous studies (\citealt{2021A&A...645A..66O, 2022A&A...666A...2O}). This resolution enables detailed modeling of small-scale structures within the ejecta (see also \citealt{2025arXiv250300130O} for the importance of spatial resolution in revealing the network of ejecta filaments), facilitating comparisons with JWST's high-angular-resolution observations.

In this paper, we analyzed two high-resolution simulations summarized in Table~\ref{Tab:model}. These include:

\begin{enumerate}
    \item Run W15-IIb-sh-HD+dec-hr, a HD simulation replicating the setup of W15-IIb-sh-HD-1eta-az from \cite{2022A&A...666A...2O} but without the feedback of particle acceleration (e.g., \citealt{2012ApJ...749..156O, 2018ApJ...852...84P}) and performed at higher resolution;
    \item Run W15-IIb-sh-MHD+dec-rl-hr, builds upon the previous simulation by incorporating the effects of an ambient magnetic field and of radiative losses; these radiative losses are calculated by accounting for deviations from electron-proton temperature equilibration and ionization equilibrium, as well as the ejecta composition derived from the SN.
    \end{enumerate}

The naming convention for the models is as follows: {\em W15} denotes the progenitor star model adopted (\citealt{1995ApJS..101..181W}); {\em IIb} indicates that the progenitor was stripped of its envelope, resulting in a SN Type IIb (\citealt{2017ApJ...842...13W}); {\em sh} indicates interaction with a circumstellar shell (\citealt{2022A&A...666A...2O}); {\em HD/MHD} specifies whether the simulation is HD or MHD; {\em dec} indicates inclusion of the Ni-bubble effect (resulting from radioactive decay chain $^{56}$Ni $\rightarrow$ $^{56}$Co $\rightarrow$ $^{56}$Fe); {\em rl} indicates the inclusion of the radiative losses from optically thin plasma; and {\em hr} identifies high-resolution simulations ($2048^3$ grid points).

Additionally, we conducted a series of test simulations, detailed in Appendix~\ref{app:losses}, to examine the effects of radiative losses and spatial resolution on the results. These simulations, performed with varied resolutions, were compared to evaluate their capacity to reproduce the small-scale structures observed in \casa. This analysis highlights the critical role of high resolution in accurately capturing the intricate ejecta features revealed by high-angular-resolution JWST observations, including also the network of unshocked ejecta filaments (see also \citealt{2025arXiv250300130O}). 

The effects of the magnetic field in simulations without radiative losses have been addressed in previous studies (\citealt{2021A&A...645A..66O, 2022A&A...666A...2O}). These works show that, in the absence of radiative cooling, the magnetic field has a negligible impact on the overall dynamics, with simulations producing qualitatively similar results whether or not the magnetic field is included. Specifically, in \cite{2022A&A...666A...2O}, we investigated the role of the magnetic field during the remnant-shell interaction and found that the parameters of the circumstellar shell required to reproduce observations remain largely unaffected by the presence of an ambient magnetic field.

\section{Results}
\label{sec:results}

The evolution of the neutrino-driven SN from core collapse to the fully developed SNR has been comprehensively detailed in previous works. The initial 20 hours, covering the expansion from core collapse well beyond the breakout of the shock wave at the stellar surface, are described in \cite{2017ApJ...842...13W}. The subsequent evolution of the blast wave through the CSM up to an age of 2000 years is presented in \cite{2021A&A...645A..66O}. For a detailed discussion of the SN-SNR evolution, we refer the reader to these studies.

\begin{table}
\caption{Setup for the simulated models.}
\label{Tab:model}
\begin{center}
\begin{tabular}{llll}
\hline
\hline
SN Model & $E\rs{exp}$ & $M\rs{ej}$ & $E\rs{exp}/M\rs{ej}$  \\
\hline
W15-2-cw-IIb$^{a}$  &  1.5 &  3.3 & 0.45  \\ 
   &  [B]$^{b}$ &  $[M_{\odot}]$ &  $[$B$/M_{\odot}]$  \\
\\ \hline\hline
SNR Model       & sim. & rad. &  radiative\\ 
                &        & decay     &  losses \\ \hline
W15-IIb-sh-HD+dec-hr     & HD    & yes     & no  \\
W15-IIb-sh-MHD+dec-rl-hr & MHD   & yes     & yes \\
\hline
\end{tabular}
\end{center}
$(a)$ Model presented in \cite{2017ApJ...842...13W};
$(b)$ Where 1 B $ = 10^{51}$~erg.
\end{table}

More recently, the SN-SNR model has been used to explore the origins of the unusual asymmetries observed in the reverse shock of \casa\ (\citealt{2022A&A...666A...2O}), including the inward or stationary motion of the reverse shock in the northwest region (\citealt{2022ApJ...929...57V, 2024ApJ...974..245S, 2025arXiv250107708F}). The study demonstrated that these asymmetries can be interpreted as signatures of the remnant's interaction with a dense, asymmetric circumstellar shell. Furthermore, the modeling provided valuable constraints on the shell's properties, linking its formation to asymmetric mass loss triggered by a massive eruption in the progenitor star approximately $10^4-10^5$ years before the SN explosion.

In this paper, we build on this latter work to investigate whether the GM observed by JWST is a relic of the shell predicted by our earlier simulations and to uncover the origin of its distinctive pockmarked structure. Our focus is on the processes responsible for forming the holes and rings within the shell, which exhibit characteristics strikingly similar to those observed in the GM. 

As we have done in previous studies (e.g., \citealt{2021A&A...645A..66O}), in the following we rotated the original simulations around the three axes by $ix = -30^\circ$, $iy = 70^\circ$, and $iz = 10^\circ$ to align the Ni-rich fingers from the SN model (\citealt{2017ApJ...842...13W}) with the Fe-rich regions observed in \casa, to facilitate comparison with observations. In this orientation, Earth’s view corresponds to the negative $y$-axis.

\subsection{Formation of holes and rings in the shocked shell}
\label{sec:shell_evol}

According to \cite{2022A&A...666A...2O}, the circumstellar shell surrounding \casa\ is relatively thin, with an initial pre-shock thickness of $\sigma \approx 0.02$~pc. The shell was located at a radius of $R_{\rm sh} \approx 1.5$~pc from the explosion center at the time of the remnant's impact. The pre-shock shell density varies between approximately $2$ and $50$~cm$^{-3}$, reflecting an asymmetric mass loss from the progenitor. Its densest portion is located on the blueshifted near side of the remnant, toward the northwest. Interestingly, this position aligns well with that inferred for the GM (\citealt{2024ApJ...976L...4D, 2024ApJ...964L..11V}), albeit slightly shifted to the north. 

The above shell position in the modeling study was constrained using the velocity measurements of the reverse shock from Chandra observations (\citealt{2022ApJ...929...57V}) and the evidence of a 3D asymmetry in \casa\ derived from Doppler velocity measurements (e.g., \citealt{1995ApJ...440..706R, 2010ApJ...725.2038D, 2013ApJ...772..134M}). In particular, the latter was crucial in determining that the shell is located on the near side of the remnant. In fact, this Doppler velocity asymmetry can be explained as the effect of a denser shell on the blueshifted near side that limits the forward expansion of the ejecta toward the observer.

As a result, our preferred model (run W15-IIb-sh-HD-1eta-az in \citealt{2022A&A...666A...2O}) assumes the shell is rotated by $50^{\circ}$ about the $z$-axis, counterclockwise from the $[x,z]$ plane, placing its densest portion on the blueshifted near side to the northwest. This configuration successfully reproduces the apparent redshift of the center of expansion, with a velocity consistent with observational estimates (\citealt{2013ApJ...772..134M}), as well as the properties of the reverse shock (\citealt{2022ApJ...929...57V}). These findings suggest that the GM, the peculiar behavior of the reverse shock in the western hemisphere, and the redshifted center of expansion in \casa\ can all be interpreted as manifestations of the remnant's interaction with an asymmetric circumstellar shell. Another sign of this interaction is the presence of circumstellar dense clumps with negligible velocity often referred to as quasi-stationary flocculi (QSFs; e.g., \citealt{1995ApJ...440..706R, 2018ApJ...866..139K}). 

   \begin{figure*}
   \centering
   \includegraphics[width=0.98\textwidth]{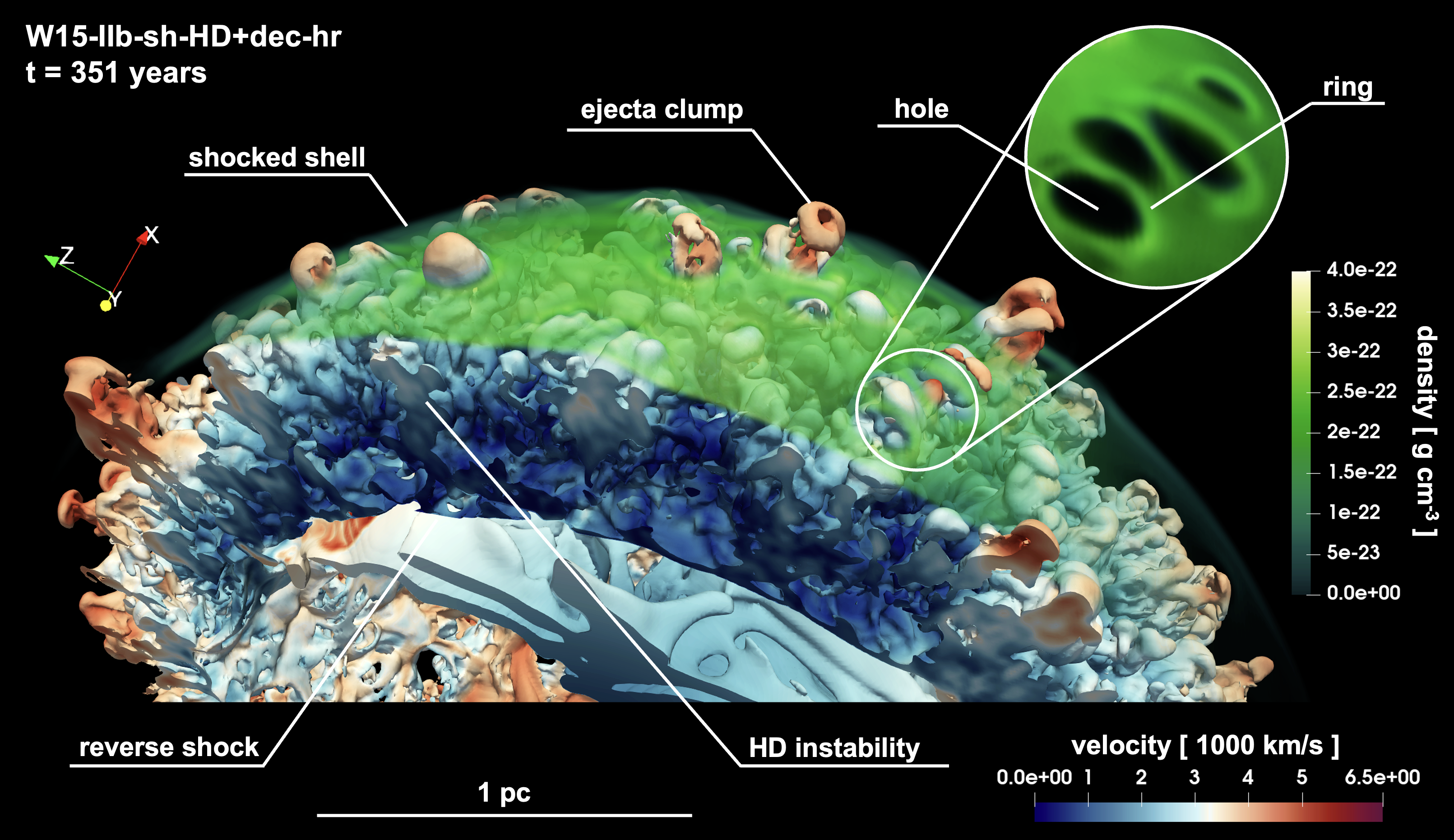}
   \caption{3D visualization of ejecta-shell interaction in \casa\ from model W15-IIb-sh-HD+dec-hr (see Table~\ref{Tab:model}). The irregular isosurface describes ejecta with mass density $> 10^{-23}$~g~cm$^{-3}$; the colors represent the radial velocity in units of 1000~km~s$^{-1}$ (the color coding is defined in the bottom right corner of the figure). The green volume rendering describes the mass density of the shocked shell material (color bar on the right of the figure). The upper right inset highlights the shell structure characterized by holes and rings, resulting from  interaction with the ejecta.}
   \label{HD-dec}%
   \end{figure*}

The overall interaction of the remnant with the shell is described in \cite{2022A&A...666A...2O} and is reproduced, at higher resolution, in Fig.~\ref{HD-dec}. More specifically, the figure provides a detailed visualization of the ejecta structures and their interactions with the shell at \casa's current evolutionary stage, derived from model W15-IIb-sh-HD+dec-hr (see Table~\ref{Tab:model}). The forward shock encountered the circumstellar shell approximately 200 years after the SN event, causing a significant deceleration as it passed through the dense material. At around 250 years post-explosion, ejecta fingers and clumps driven by HD instabilities and emerging from the contact discontinuity began penetrating the shocked shell material. This mechanism led to distinctive morphological features characterized by holes generated by the ejecta within the shell structure (see also \citealt{2024ApJ...976L...4D}). Figure~\ref{HD-dec} illustrates the basic mechanism underlying hole formation and the complex interplay between ejecta dynamics and shell structure. As the ejecta structures propagated through the shell, they displaced the surrounding shell material, which subsequently accumulated along the peripheries of these penetration sites. This process produced dense, ring-like structures encircling the newly formed holes. The progressive evolution of these interactions resulted in a characteristic pockmarked pattern analogous to that observed in \casa\ today.

Although the general morphology of holes and rings is very similar to that observed in \casa, we found that most of the simulated features have diameters in the range of $\approx 5^{\prime\prime} - 20^{\prime\prime}$ (corresponding to $\sim 0.07-0.3$~pc). As noted by \cite{2024ApJ...976L...4D}, these dimensions are significantly larger than those observed in the GM, which range from $\approx 1^{\prime\prime}- 3^{\prime\prime}$. This discrepancy raises some questions on the interpretation of the GM's structure based on the mechanism described above. The same authors suggested that the size difference may arise from the limited spatial resolution of the simulations considered, which hinders an accurate representation of the small-scale structure of the ejecta. However, as shown in Fig.~\ref{HD-dec} for run W15-IIb-sh-HD+dec-hr (see Table~\ref{Tab:model}), our analysis revealed that the diameters of the holes and rings do not change significantly compared to earlier, lower-resolution simulations (presented in \citealt{2022A&A...666A...2O}), indicating that the spatial resolution was sufficient to capture the size of the holes for the specific modeled interaction analyzed.

   \begin{figure*}
   \centering
   \includegraphics[width=0.98\textwidth]{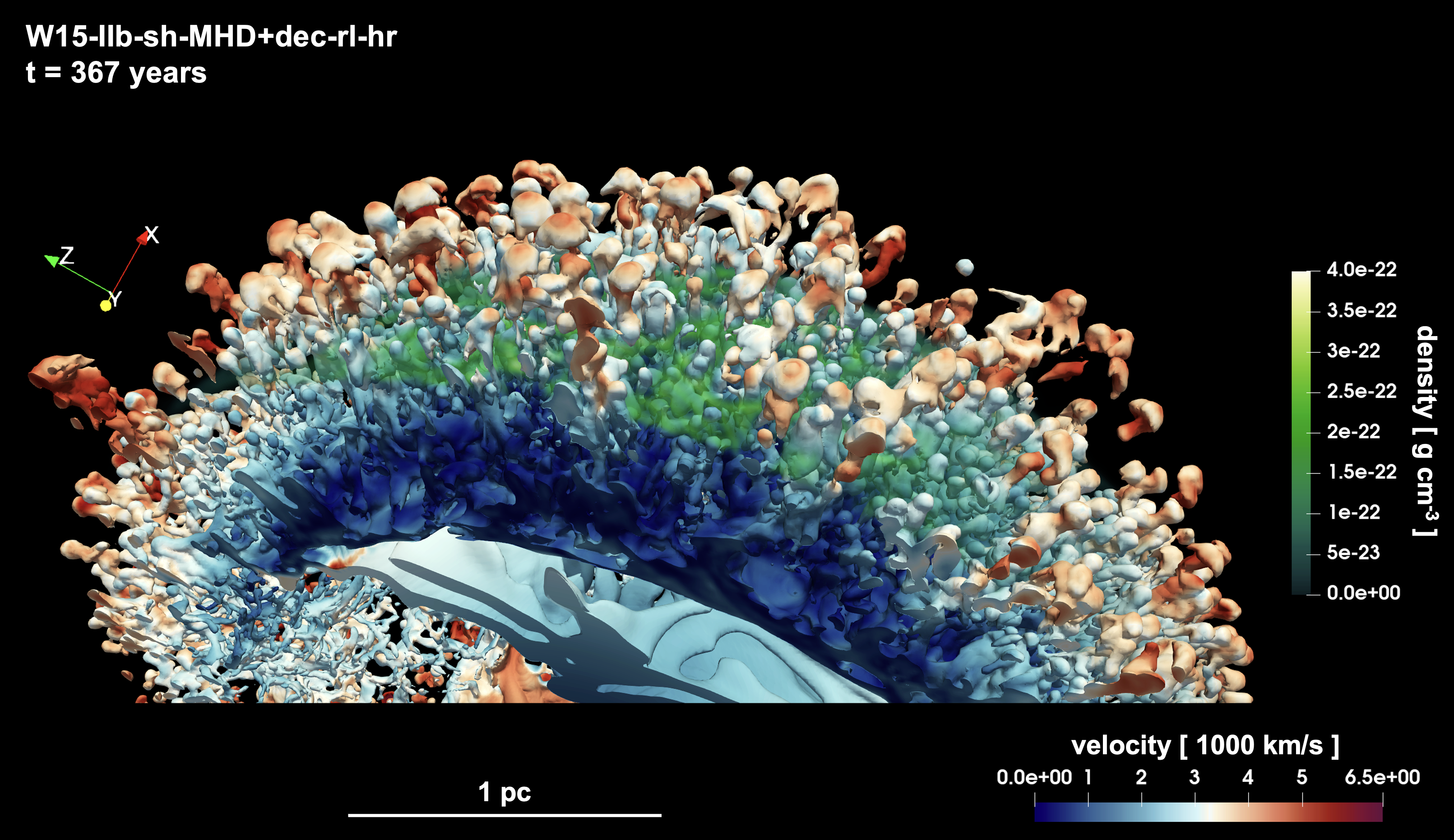}
   \caption{Same as Fig.~\ref{HD-dec}, but for the model W15-IIb-sh-MHD+dec-rl-hr, which includes the effects of radiative cooling and magnetic fields. See online Movie~1 for an animation of these data. A navigable 3D graphic of the ejecta and shell structures at the age of \casa\ is available at https://skfb.ly/psYpK.}
   \label{MHD-dec-rl}%
   \end{figure*}

This result indicates that additional factors may contribute to the observed discrepancies between the models and observations. To address this, we investigated the impact of incorporating more detailed microphysical processes into the modeling, such as radiative cooling and effects of magnetic fields. In fact these processes can lead to smaller ejecta structures by enhancing their compression through radiative losses and magnetic confinement. To explore this, we conducted run W15-IIb-sh-MHD+dec-rl-hr, an MHD simulation using the same configuration as W15-IIb-sh-HD+dec-hr but incorporating the effects of radiative cooling and magnetic fields (see Table~\ref{Tab:model}). By comparing the two simulations, we investigated how these additional processes influence the dynamics and morphology of the ejecta and the shocked shell.

In run W15-IIb-sh-MHD+dec-rl-hr, we found that radiative cooling becomes efficient in the mixing region between the reverse and forward shocks approximately 10 years after the core collapse, well before the plumes of ejecta enriched in heavy elements from the SN explosion begin interacting with the reverse shock (\citealt{2021A&A...645A..66O}). The cooling efficiency is particularly pronounced in regions rich in heavier elements, driven by strong line emission from ions such as Fe, Si, and O. Around 30 years after the explosion, as Fe-rich plumes begin to interact with the reverse shock, radiative losses intensify significantly, especially in the newly formed shocked Fe-rich regions. Additional test simulations discussed in Appendix~\ref{app:losses} highlight the importance of incorporating non-equilibrium of ionization (NEI) effects into radiative cooling calculations. Specifically, we found that freshly shocked plasma, initially underionized due to NEI effects, exhibits significantly enhanced radiative losses via strong emission lines compared to plasma in collisional ionization equilibrium (CIE) at the temperatures typical of the mixing region. These NEI effects are therefore crucial for amplifying radiative cooling and, consequently, for accurately modeling the structure of shocked ejecta.

Indeed we found that radiative cooling plays a pivotal role in shaping the ejecta fingers formed by HD instabilities at the contact discontinuity. For comparison, Fig.~\ref{MHD-dec-rl} shows the structure of shocked ejecta at the age of \casa\ that appears more fragmented and complex than the structure found in run W15-IIb-sh-HD+dec-hr (Fig.~\ref{HD-dec}). The evolution of these data from $\approx 200$~year after the core collapse to the age of $\approx 1000$~yr is provided as online Movie~1. Cooling, by converting thermal energy into radiation, reduces the post-shock temperature, resulting in decreased pressure and increased compression. This process increases the density of the post-shock material, further enhancing the efficiency of radiative cooling and facilitating the formation of dense knots and filamentary structures visible in Fig.~\ref{MHD-dec-rl} (see also the online Movie~1).

   \begin{figure}
   \centering
   \includegraphics[width=0.46\textwidth]{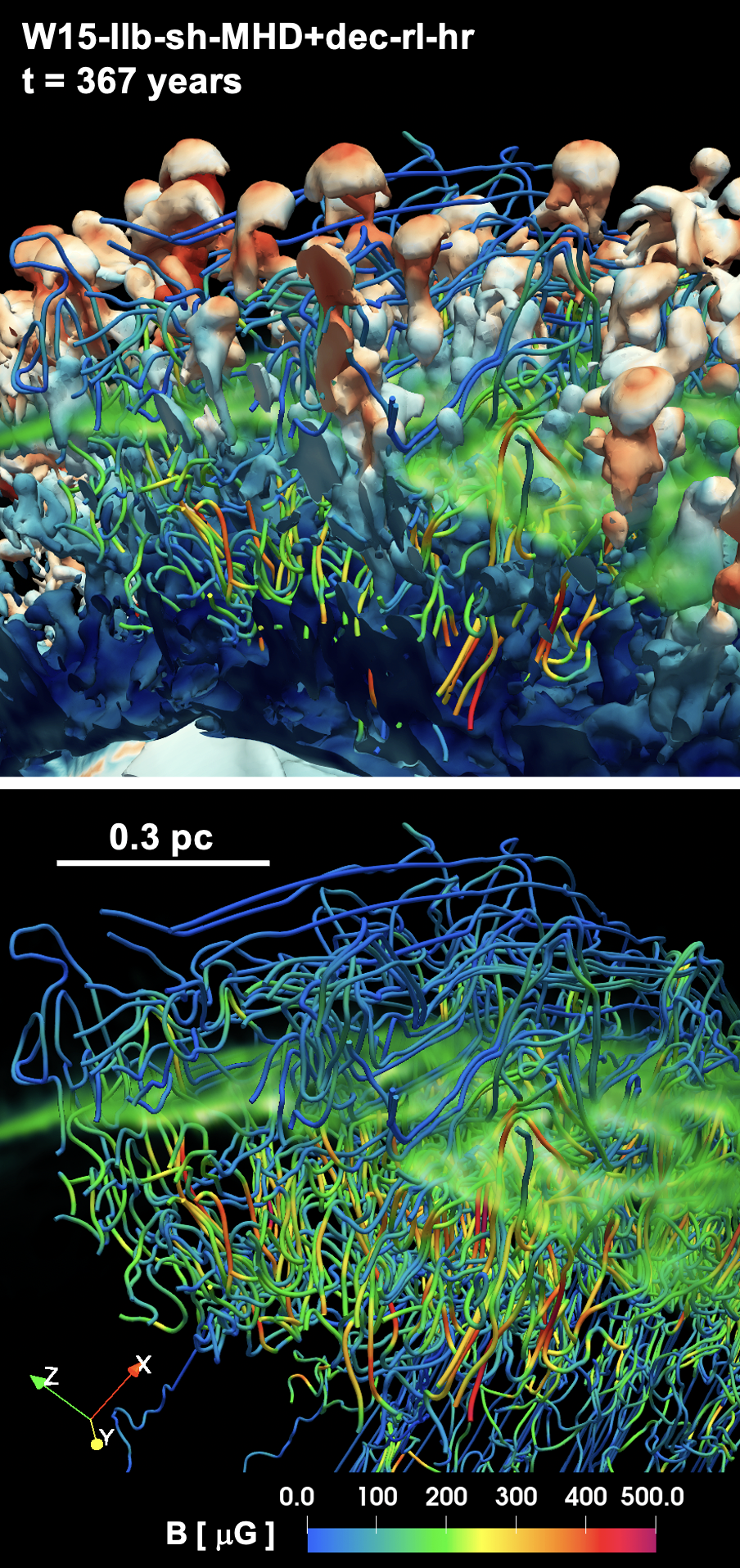}
   \caption{Close-up view of the ejecta fingers and clumps emerging from the contact discontinuity, along with the perturbed magnetic field, in model W15-IIb-sh-MHD+dec-rl-hr at the age of \casa. The irregular isosurface represents the ejecta, while the green volumetric rendering shows the shocked shell, as described in Fig.~\ref{HD-dec}. The sampled magnetic field lines illustrate the magnetic field configuration, with colors indicating the magnetic field strength in units of $\mu$G (color bar at the bottom of the figure). The lower panel is the same as the upper panel, but with the isosurface removed to provide a clearer view of the complex magnetic field configuration. A navigable 3D graphic of the magnetic field configuration at the age of \casa\ is available at https://skfb.ly/psYpK.}
   \label{bfield}%
   \end{figure}

The magnetic field adds an important ingredient in this context (see Fig.~\ref{bfield} and the online Movie~1). After interaction with the forward shock, it assumes a rather complex topology due to strong perturbations by the ejecta fingers and clumps emanating from the contact discontinuity (see the deformed magnetic field lines in the figure). Although it does not substantially affect the overall dynamics of the remnant, the field plays a significant role in stabilizing smaller-scale structures (see also Appendix~\ref{app:losses}). Specifically, it partially suppresses HD instabilities at the boundaries of clumps and ejecta fingers, thereby preserving these features and reducing the probability of their rapid fragmentation (e.g., \citealt{2012ApJ...749..156O}). In some cases, sufficiently dense or fast-moving ejecta fragments (or bullets) can survive HD instabilities and reach the outer layers of the remnant, sometimes even protruding beyond its outline as fast shrapnels (e.g., \citealt{2001ApJ...559L..45M}). The survival of these fragments also depends on their density contrast, velocity contrast, and the surrounding medium’s conditions, which determine their ability to resist disruption in the mixing region (e.g., \citealt{2002ApJ...574..155W, 2013MNRAS.430.2864M, 2015MNRAS.453..166T}). Additionally, the magnetic field enhances the cooling efficiency by confining the plasma, increasing its density, and amplifying the effects of radiative losses (e.g., \citealt{2008ApJ...678..274O}). According to our simulation, the perturbed magnetic field strength can reach values in the mixing region as high as a few milligauss (see Fig.~\ref{bfield}), consistent with estimates derived from infrared observations collected using the Stratospheric Observatory for Infrared Astronomy (SOFIA; \citealt{2023MNRAS.522.2279R}).

   \begin{figure*}
   \centering
   \includegraphics[width=0.84\textwidth]{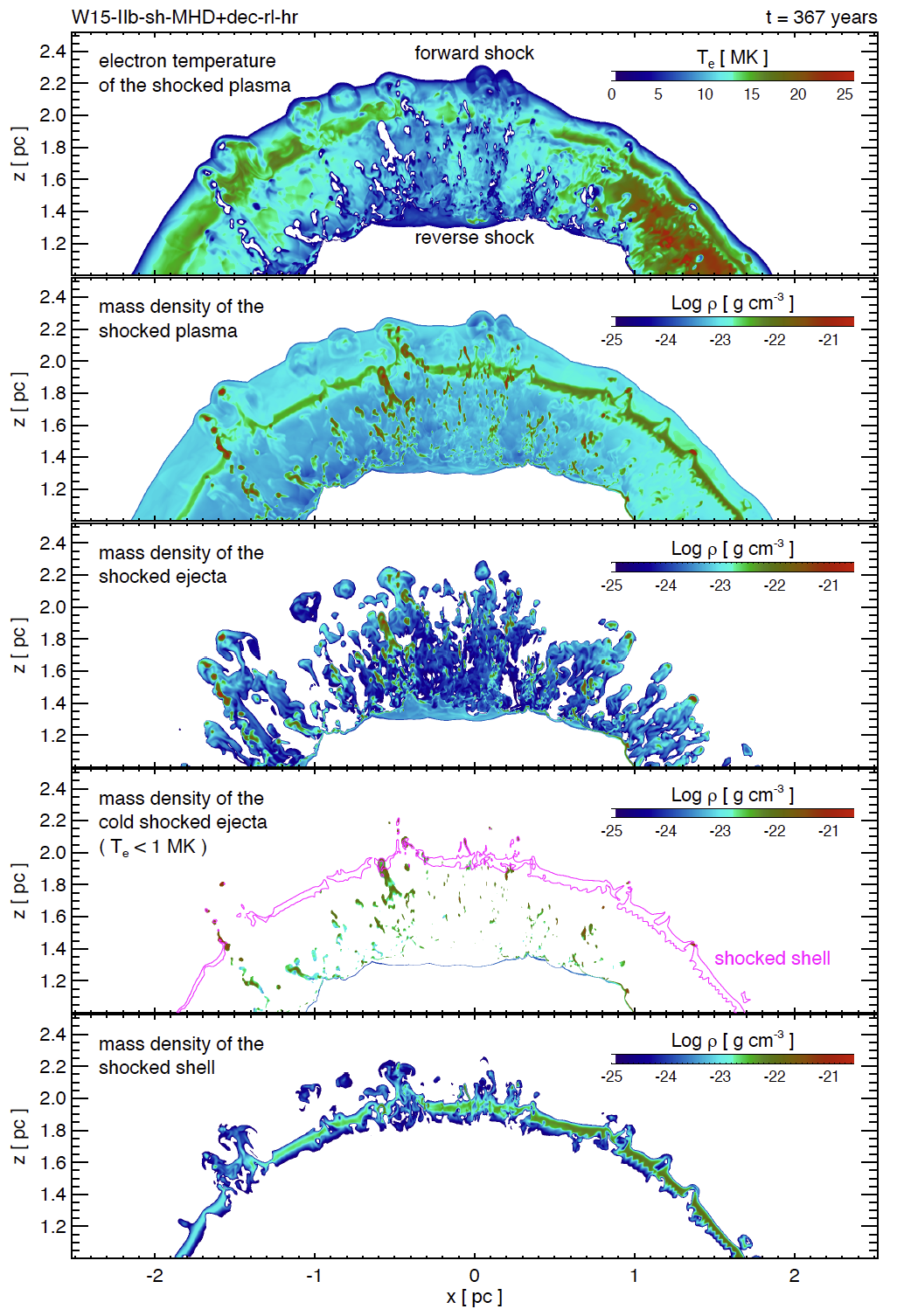}
   \caption{2D slices in the $(x, z)$ plane showing the spatial distributions of temperature and density for shocked plasma between forward and reverse shock at the age of \casa, based on model W15-IIb-sh-MHD+dec-rl-hr. The panels display (from top to bottom): electron temperature, mass density of shocked plasma (CSM and ejecta), shocked ejecta density, shocked ejecta density with electron temperatures below 1 MK, and shocked shell density. The magenta contour in the second panel from the bottom indicates the shocked shell.}
   \label{distrib_temperature}%
   \end{figure*}

   \begin{figure*}
   \centering
   \includegraphics[width=\textwidth]{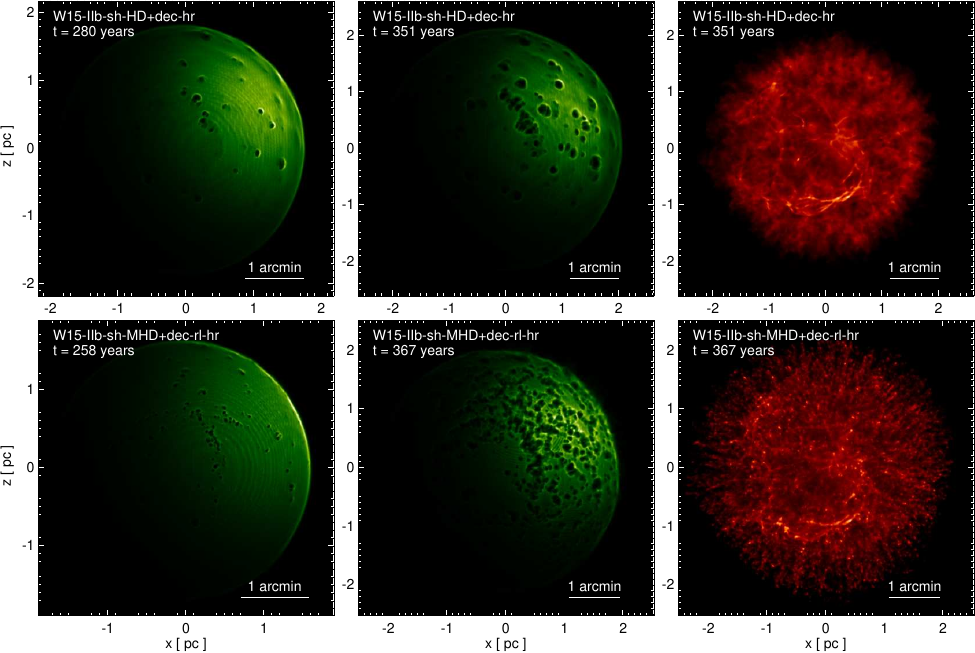}
   \caption{The green volumetric renderings in the first two columns represent the shocked circumstellar shell as derived from models W15-IIb-sh-HD+dec-hr (upper panels) and W15-IIb-sh-MHD+dec-rl-hr (lower panels) at two evolutionary stages: approximately 100 years before the age of \casa\ (left column) and at the age of \casa\ (center column). The red volumetric rendering in the last column illustrates the corresponding ejecta distributions at the age of \casa.}
   \label{evol_gm}%
   \end{figure*}

The radiative cooling has a strong impact on the temperature structure of the shocked plasma in the mixing region between the forward and reverse shocks. Figure~\ref{distrib_temperature} presents 2D slices in the $(x,z)$ plane of the northern hemisphere, illustrating the spatial distributions of electron temperature (top panel) and mass density (remaining panels) of the shocked plasma on a logarithmic scale, as modeled in W15-IIb-sh-MHD+dec-rl-hr. The 2D slices cut through the circumstellar shell. The bottom panel highlights a cross-section of the shocked shell, with its densest region oriented in the positive $x$-axis. For easier comparison, the second panel from the bottom includes contours outlining the shell material, allowing for a clearer correlation between the shell's morphology and the positions of dense, cold ejecta knots.

The electron temperature peaks at approximately 20 MK, primarily in regions where the remnant interacts with the densest parts of the shell. This high temperature is due to additional heating from the reflected shock generated during the interaction. In contrast, the shocked ejecta exhibit dense, compact knots with electron temperatures below 1 MK, reaching as low as tens of kelvin (see second panel from the bottom in the figure). These cold knots are the result of significant radiative cooling, which triggers thermal instabilities that collapse fragments of ejecta. This process predominantly occurs at the heads of Rayleigh-Taylor fingers forming at the contact discontinuity. These dense, cold, and compact knots behave like bullets, penetrating through the overlying ejecta layers. They are among the first structures to break through the circumstellar shell, forming the observed holes (as illustrated in the lower two panels of Fig.\ref{distrib_temperature}). In some cases, they even overcome the forward shock, creating prominent protrusions visible in the two upper panels of the figure. 

Figure~\ref{distrib_temperature} shows that, by the age of \casa, the structure of the mixing region is significantly transformed when radiative losses and magnetic fields are incorporated into the simulations (compare also Figs.~\ref{HD-dec} and \ref{MHD-dec-rl}). This region becomes dominated by thin, highly fragmented streams of shocked ejecta that expand forward more efficiently than in the absence of radiative cooling and magnetic confinement. As a result, the inclusion of these processes allows ejecta clumps to extend to greater radial distances and enhances the mixing of ejecta with the CSM (compare the extension of ejecta fingers in Figs.~\ref{HD-dec} and \ref{MHD-dec-rl}). The increased fragmentation of the ejecta is particularly impactful, as it leads to a heavily pockmarked structure in the shocked shell. Numerous small-scale ejecta fragments create a diverse array of holes and rings of varying sizes. Our simulations indicate that the diameters of these features range from $1^{\prime\prime}$ to $12^{\prime\prime}$, with the majority of holes having diameters of approximately $5^{\prime\prime}$. While this range is broadly consistent with the sizes of holes observed in the GM, the simulations also produce a notable number of holes with diameters larger than those observed. This discrepancy highlights limitations in the modeled size distribution, suggesting areas for further refinement in future simulations, as discussed in the next section.

\subsection{Properties of the holes and rings over time and comparison with JWST observations}
\label{sec:holes_rings}

The two simulations discussed above reveal that clumps and fingers of ejecta begin to penetrate the shell approximately 100 years before the current age of \casa\ (see the online Movie~1). During this phase, a few sparse holes start to appear on the previously smooth surface of the shocked shell. Figure~\ref{evol_gm} shows the shell's structure at two stages: when the first holes form ($\approx 250$ years after the explosion) and at the age of \casa. In the latter case, the figure also shows the corresponding ejecta distribution. In run W15-IIb-sh-MHD+dec-rl-hr, the holes form earlier and are relatively smaller compared to those in run W15-IIb-sh-HD+dec-hr. This difference arises because, in the former simulation, the ejecta fingers are thinner (producing smaller holes) and more elongated outward (starting the interaction with the shell earlier) due to the combined effects of radiative cooling and magnetic confinement, which are absent in the latter simulation. Additionally, the number of holes is greater in run W15-IIb-sh-MHD+dec-rl-hr, reflecting the increased fragmentation of the ejecta fingers induced by these effects.

   \begin{figure*}
   \centering
   \includegraphics[width=0.98\textwidth]{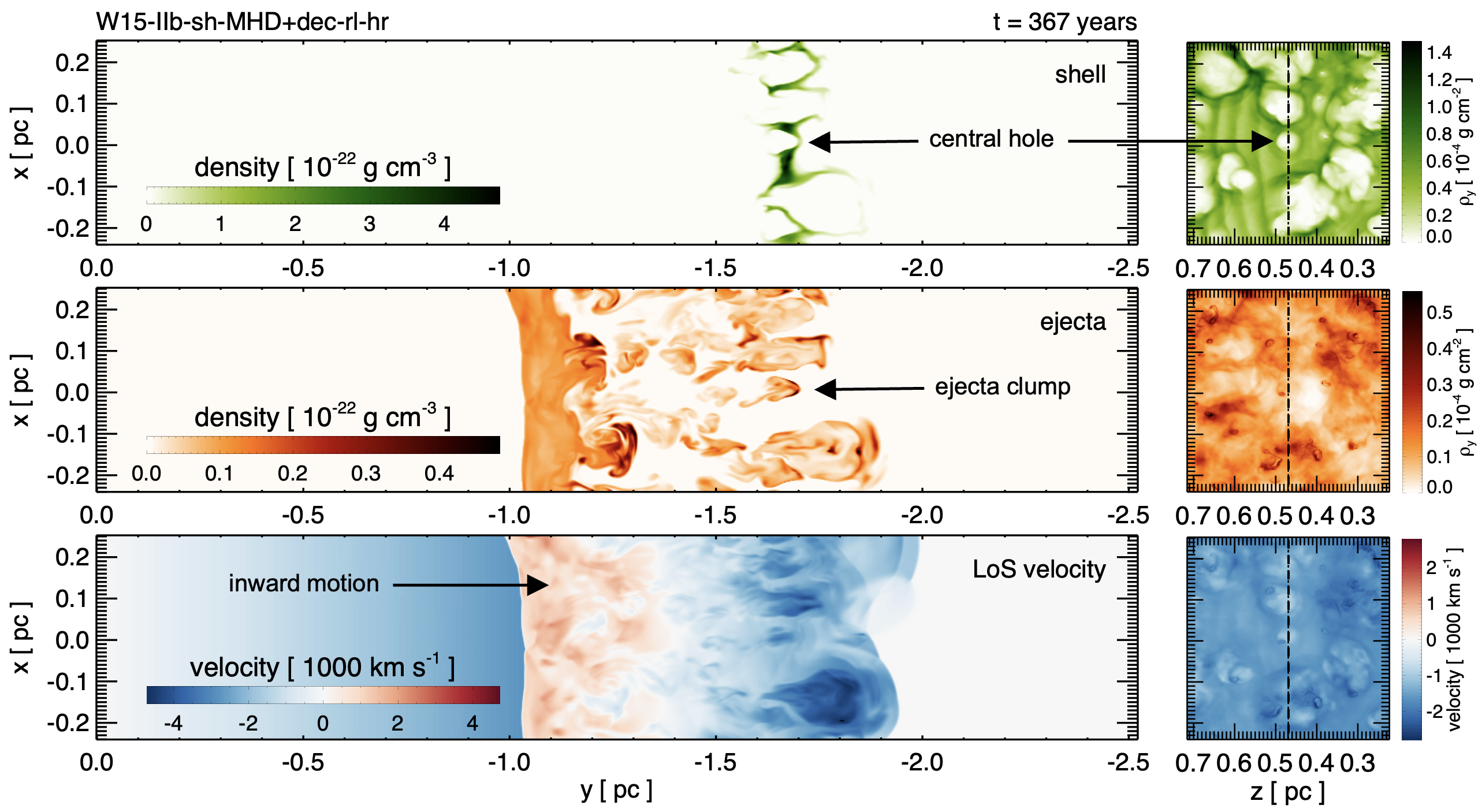}
   \caption{Visualization of the plasma structure in a region where the SN ejecta interact with the circumstellar shell in model W15-IIb-sh-MHD+dec-rl-hr at the age of \casa. This region corresponds to an enlarged view of region~6 in Fig.~\ref{comparison}, highlighted by the dashed box. The left panels display cross-sections of the shell density (top), shocked ejecta density (middle), and LoS velocity (bottom) in a plane oriented along the LoS. The density distribution of the unshocked ejecta in the interior is not shown. The right panels provide 3D volumetric renderings of the same quantities within the selected region, as seen from Earth's vantage point: surface density of the shell (top) and of the shocked ejecta (middle), and the density weighted LoS velocity average (bottom). The vertical dashed line in the 3D renderings marks the location of the cross-sections shown in the left panels, facilitating direct comparison between the two perspectives. See online Movie~2 for an animation of these data. A navigable 3D graphic of the evolution of this region is available at https://skfb.ly/ptFHY.}
   \label{holes_struct}%
   \end{figure*}

The shocked shell exhibits densities in the range between 30 and $150$~cm$^{-3}$ in the dense near side, which is roughly consistent with density values inferred from ionization age estimates based on X-ray observations ($\approx 48$~cm$^{-3}$ corresponding to a pre-shock density of $\approx 12$~cm$^{-3}$; \citealt{2024ApJ...964L..11V}). The thickness of the shocked shell is reduced relative to its pre-shock value ($0.02$~pc) due to shock compression and is approximately $\sim 0.015$~pc in run W15-IIb-sh-HD+dec-hr and $\sim 0.007$~pc in run W15-IIb-sh-MHD+dec-rl-hr. In the latter case, this further reduction is driven by the enhanced compression resulting from radiative cooling. The shell's thickness in both simulations is larger than the value inferred from JWST observations ($\sim 0.002$~pc; \citealt{2024ApJ...976L...4D}).

It is worth noting, however, that the thickness derived from observations may be influenced by the dominance of dust emission in the MIRI images. The inferred shell thickness could be affected by the lifetime of dust grains, which are susceptible to sputtering in the shocked gas. For instance, \cite{2024ApJ...976L...4D} estimated a sputtering timescale of 33 years for a specific density and grain size. Consequently, the actual shell thickness could be larger than the observed value. Additionally, in our simulations, the shell thickness and the size of the smallest holes produced are comparable to the spatial resolution. For example, in run W15-IIb-sh-MHD+dec-rl-hr, the shell thickness is represented by only about five grid points. This suggests that, despite the high resolution, the simulations may not fully resolve dynamics at such small scales, which could contribute to discrepancies with observational measurements.

Figure~\ref{holes_struct} illustrates the plasma structure in a region where the ejecta interact with the circumstellar shell in model W15-IIb-sh-MHD+dec-rl-hr (see Movie~2 for the evolution of this region). This region is located in projection near the center of the remnant, causing the holes to appear almost pole-on from the Earth's vantage point. The left panels show cross-sections of density and velocity in a plane oriented along the line-of-sight (LoS), perpendicular to the plane of the sky, offering insights into the radial properties of the ejecta clumps and perturbed shell structure. The right panels display 3D volumetric renderings of the same quantities, as seen from Earth's vantage point to replicate observational conditions similar to those of the GM. 

The example shown in the figure represents a region located in projection near the center of the remnant, where the shell exhibits higher densities. The central hole in the field of view (see upper right panel in Fig.~\ref{holes_struct}) has a diameter of $\approx 0.05$~pc (corresponding to $\approx 3^{\prime\prime}$ at the distance of \casa, consistent with the maximum dimensions of holes inferred from JWST observations). The density at the border of the hole is $\rho \approx 4\times 10^{-22}$~g~cm$^{-3}$ (particle number density $\approx 180$~cm$^{-3}$), forming a ring that is denser than the surrounding shocked shell by approximately a factor of two. The ejecta structure in this region is highly complex and turbulent, as illustrated in the middle panels (see also Movie~2). A compact clump of ejecta responsible for the formation of the central hole is visible in the middle left panel. This clump has a density approximately an order of magnitude lower than that of the ring ($\rho \approx 0.4 \times 10^{-22}$~g~cm$^{-3}$) but is able to penetrate the shell due to its high blueshifted velocity ($\approx -3500$~km~s$^{-1}$). 

\begin{figure}
   \centering
   \includegraphics[width=0.45\textwidth]{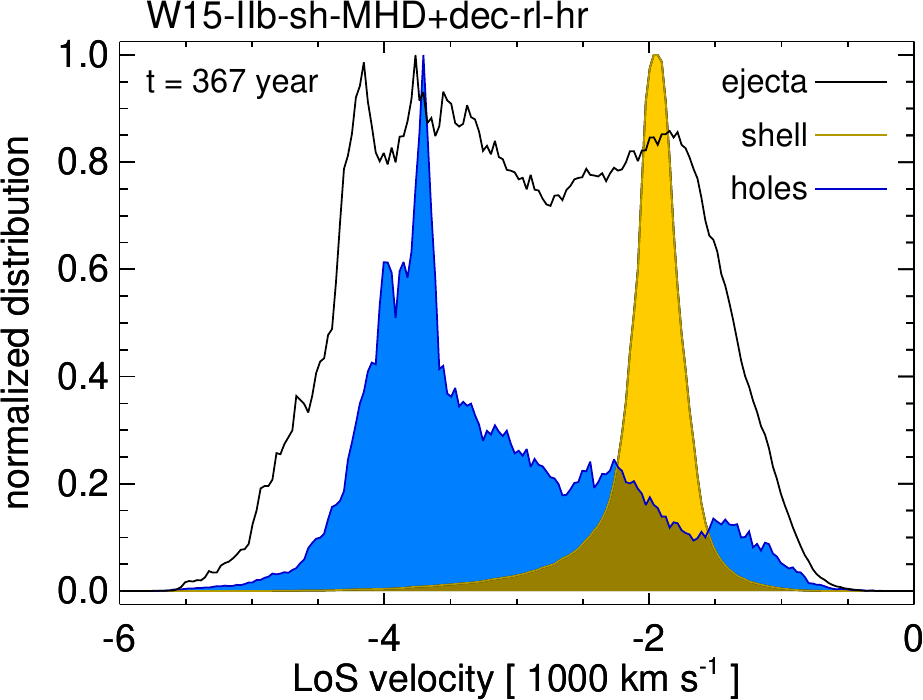}
   \caption{Normalized LoS velocity distribution of the shocked ejecta (black curve), shocked shell (orange curve), and the shocked ejecta filling the holes (blue curve) in the region shown in Fig.~\ref{holes_struct}. The analysis is restricted to plasma with coordinates $y<-1.5$~pc, corresponding to the domain where the shocked shell interacts with the ejecta. The results are derived from the W15-IIb-sh-MHD+dec-rl-hr model at the evolutionary stage corresponding to \casa.}
   \label{distrib_velocity}%
   \end{figure}

The bottom panels of Fig.~\ref{holes_struct} highlight an inward motion of the plasma (in red) below the shocked shell caused by the reflected shock generated by the interaction between the ejecta and the shell (see also Movie~2). This reflected shock is responsible for the inward motion of the reverse shock in this portion of the remnant, as discussed in \cite{2022A&A...666A...2O}. The shell material exhibits a LoS velocity of approximately $-2000$~km~s$^{-1}$, while the shocked ejecta display velocities ranging from $-1000$ to $-5500$~km~s$^{-1}$ (see yellow and black lines, respectively, in Fig.~\ref{distrib_velocity}). The blue curve in the figure represents the ejecta confined to the projected area of the holes, excluding those that, in projection, fall outside this area. Observations limited to the area of the holes capture only these ejecta, providing a focused view of their velocity distribution and dynamics as shown in the blue curve of Fig.~\ref{distrib_velocity}.

Interestingly, the extended ejecta fingers, which can be detected in the holes, display velocities concentrated around $-4000$~km~s$^{-1}$ (as indicated by the blue curve in Fig.\ref{distrib_velocity}). This contrasts with the presence of ejecta in the volume analyzed with velocities reaching up to $-5500$~km~s$^{-1}$ (black curve in the figure). This apparent discrepancy arises because ejecta clumps, after protruding through the shell, tend to expand, with parts of these clumps (typically their leading edges) falling outside the projected area of the holes (see Fig.~\ref{holes_struct}). Moreover, the selected region has a positive $z$-coordinate, indicating an angle between the LoS and the normal to the shell, a typical condition in observations. Consequently, the fastest ejecta, which expand radially and form an angle with the LoS, often move beyond the projected area of the holes after emerging from the shell. These high-velocity ejecta are excluded from the blue curve but are included in the black curve, as shown in Fig.~\ref{distrib_velocity}. The peak around $-4000$~km~s$^{-1}$ in the blue curve, therefore, represents the LoS velocity of ejecta that have just passed through the holes.

Since, in the analyzed region, the LoS velocity is similar to the radial velocity (the angle between the LoS and the normal to the shell is small), the velocity of the shocked shell material, including the rings, shows good agreement with the radial velocities of the GM inferred from X-ray observations ($-2300$~km~s$^{-1}$; \citealt{2024ApJ...964L..11V}). However, we also note that the models do not predict the lower radial velocities ($-50$~km~s$^{-1}$) inferred from NIRSpec and MIRI MRS line emission associated with the rings (\citealt{2024ApJ...976L...4D}). Reproducing these lower velocities would require significantly over-dense material within the asymmetric shell shaped by mass loss. This discrepancy suggests a highly inhomogeneous, clumpy CSM, through which the forward shock propagates. A more intricately structured, density-varied shell may better capture the observed velocity range.

   \begin{figure*}
   \centering  
   \includegraphics[width=0.82\textwidth]{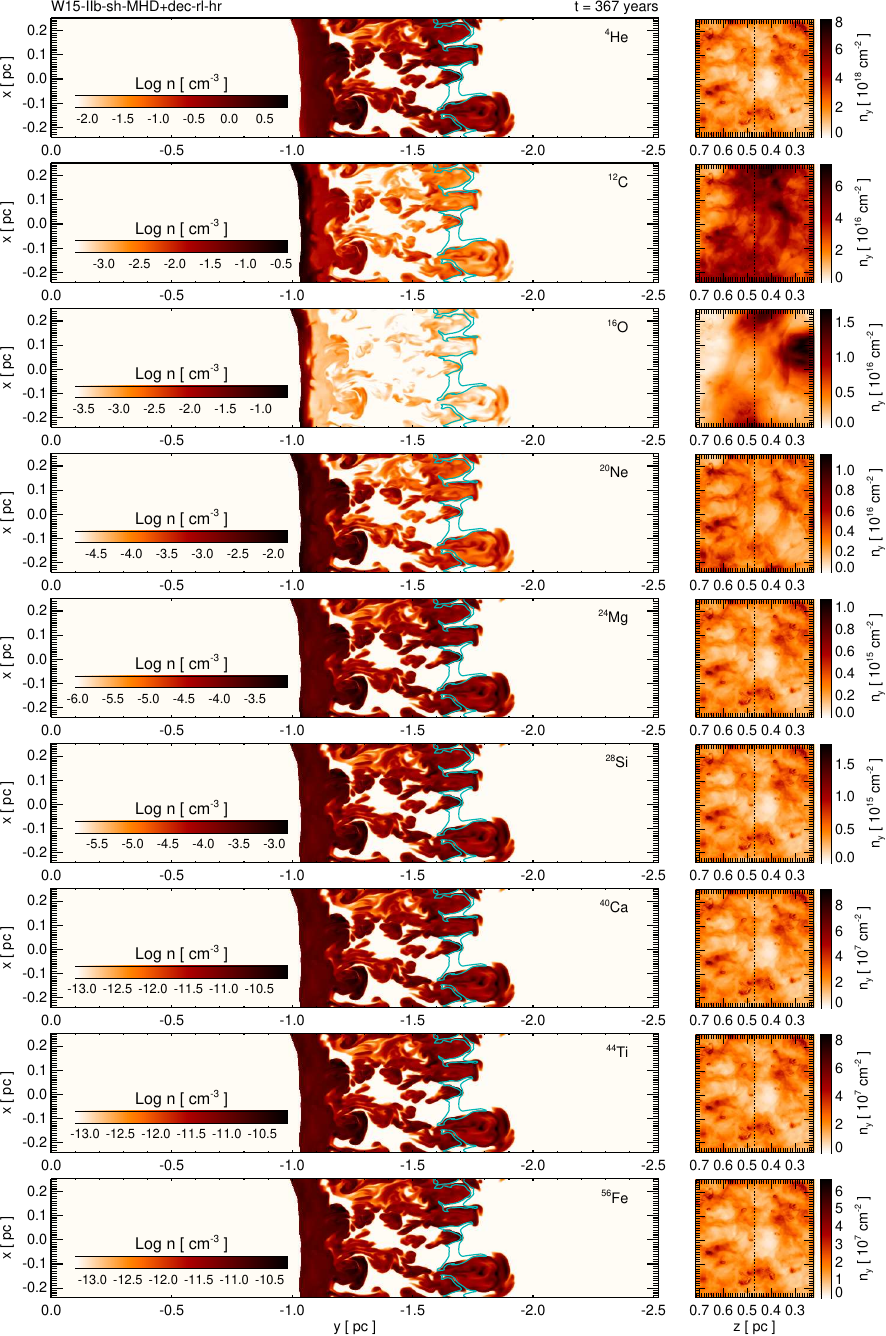}
   \caption{Left panels: Cross-sections of the density distribution for selected ejecta species in a plane aligned with the LoS, corresponding to the region shown in Fig.~\ref{holes_struct}. The color bar in each panel represents the density in log scale. The light blue contour outlines the material of the shocked shell. Right panels: Corresponding 3D volumetric renderings of the same quantities as seen from Earth's vantage point. The color bar on the right of each panel shows the column density in linear scale. The vertical dashed line marks the location of the cross-sections shown in the left panels. The results are derived from the W15-IIb-sh-MHD+dec-rl-hr model at the evolutionary stage corresponding to \casa.}
   \label{composition}%
   \end{figure*}

The density distribution of elements enriching the ejecta clumps that pierce the shell in the region analyzed is shown in Fig.~\ref{composition}. These elements include light species (He), intermediate-mass species (C, O, Ne, Mg), and heavy species (Si, Ca, Fe). In this region, the ejecta clumps are predominantly enriched in light and intermediate-mass elements, particularly He, C, O, and Ne while being relatively deficient in heavier elements. However, it is important to note that the chemical composition of the ejecta clumps in our models varies significantly across different regions of the remnant. The selected region, located near the center of the remnant’s projected image, is representative of the GM, which is also observed in projection near the remnant's center and is distinct from the Fe-rich regions that characterize the main shell of \casa. In this case, no clumps with a significant amount of Fe are present in the selected region. In contrast, in correspondence of the three Fe-rich regions present in the model (see \citealt{2021A&A...645A..66O}), the clumps and fingers protruding through the shell are notably enriched in Fe and other heavy elements. Regardless of the region, the gas filling the holes consistently exhibits higher elemental abundances compared to the surrounding shell material.

   \begin{figure}
   \centering
   \includegraphics[width=0.48\textwidth]{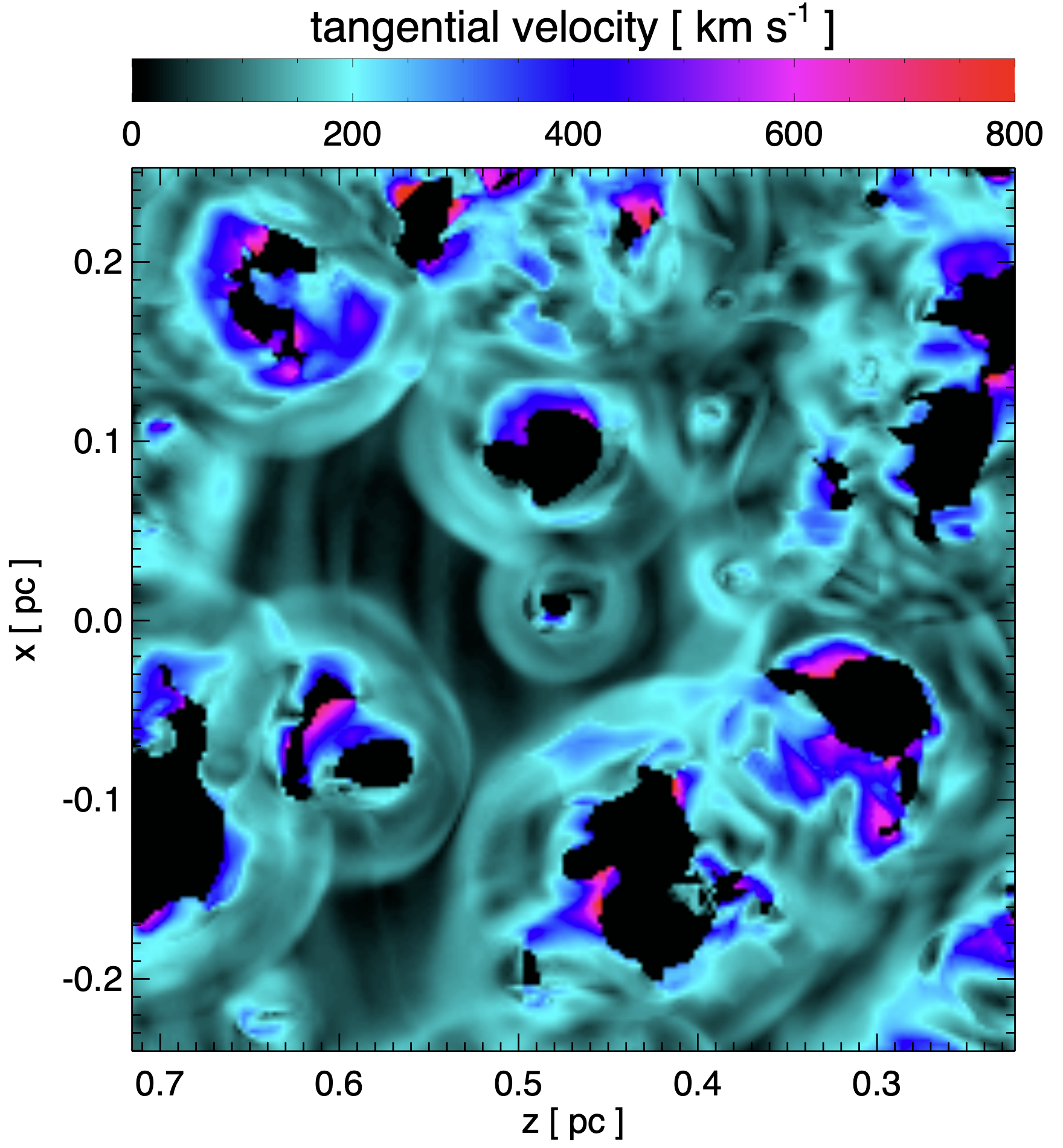}
   \caption{Tangential velocities of the shell material relative to the LoS for the region analyzed in Fig.~\ref{holes_struct}. The velocities are averaged along the LoS and weighted by the shell's density, including only cells where the density exceeds 5\% of the maximum value of the shell density in the volume analyzed. The results are derived from the W15-IIb-sh-MHD+dec-rl-hr model at the evolutionary stage corresponding to \casa. The irregular shapes of many holes are partly due to difficulty in identifying shell material and distinguishing it from ejecta within the holes (see text) and partly due to the superposition of multiple holes.}
   \label{tangent_vel}%
   \end{figure}

Another distinctive feature of the holes and rings produced by the ejecta is the potential presence of tangential velocities around the holes. According to the model, as the ejecta clumps and fingers penetrate the shell, they displace its material and generate compression waves that propagate tangentially to their radial motion. This leads to another intriguing property that can characterize the holes and ring-like features created by the ejecta fingers: the tangential motion of plasma around the holes in the plane of the sky. Figure~\ref{tangent_vel} illustrates the magnitude of the tangential velocity derived for the region analyzed in Fig.~\ref{holes_struct}. These velocities were averaged along the LoS and weighted by the shell's density, considering only numerical cells where the density exceeds 5\% of the maximum value of shell density in the volume analyzed. The black areas in the figure roughly correspond to the holes created by the ejecta clumps. Note that the irregular shapes of many holes are partly due to the challenge of distinguishing shell material from ejecta within the holes, and partly due to the superposition of multiple holes, which results in non-circular shapes. Surrounding these holes, the compression waves caused by the interaction are clearly visible, with moderate velocities of approximately 200 km~s$^{-1}$. Near the edges of the holes, the simulation reveals higher tangential velocities, ranging from 400 to 800 km~s$^{-1}$, indicating freshly displaced shell material caused by the penetrating ejecta fingers. 

Based on these results, we propose that tangential velocity measurements, derived, for example, from high-resolution observations taken at different epochs, could help distinguish between the two scenarios proposed by \cite{2024ApJ...976L...4D}. The first scenario, explored in this study, involves the interaction of the GM with ejecta clumps and fingers after the passage of the forward shock. The second scenario suggests that small, high-velocity ejecta bullets impact the GM before the arrival of the forward shock. Unlike the first scenario, the second predicts that the material surrounding the holes would stop expanding after the forward shock's impact and gradually begin to shrink (e.g., \citealt{2024ApJ...976L...4D}).

   \begin{figure*}
   \centering
   \includegraphics[width=0.99\textwidth]{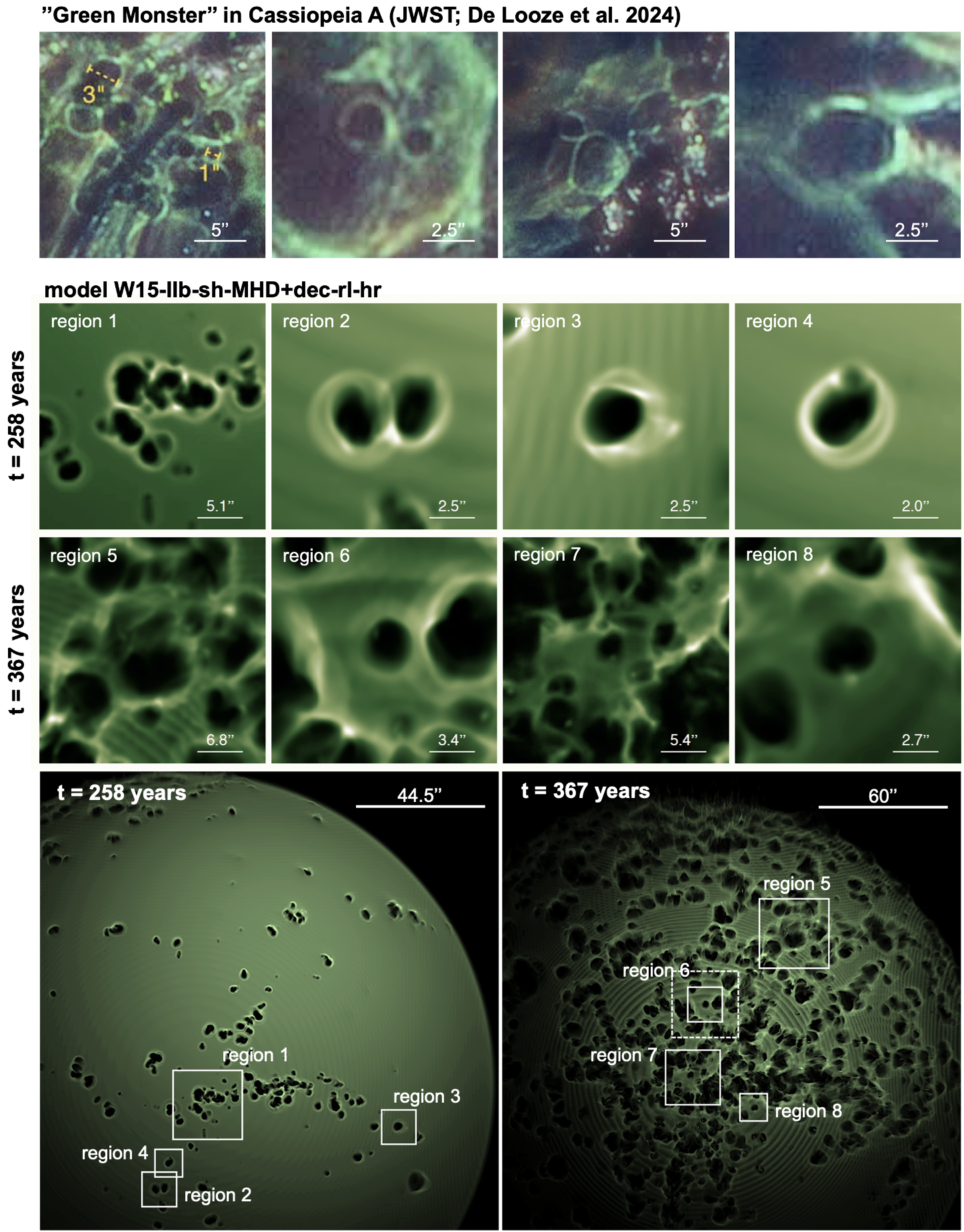}
   \caption{Examples of hole and ring structures observed in the GM with JWST (\citealt{2024ApJ...976L...4D}; first row) compared to those produced in run W15-IIb-sh-MHD+dec-rl-hr at two different epochs: 258 years (second row) and 367 years (third row). The latter epoch corresponds approximately to the current age of \casa. The bottom panels highlight the selected regions within the context of the entire shocked shell; the dashed box indicates the region analyzed in Figs.~\ref{holes_struct}, \ref{composition}, and~\ref{tangent_vel}. Hole dimensions are shown in arcseconds, assuming a distance of $3.4$~kpc to \casa.}
   \label{comparison}%
   \end{figure*}

A few examples of hole and ring morphologies identified in our simulations are presented in Fig.~\ref{comparison} and are compared to similar features observed by JWST in \casa. While some morphologies in the simulations closely resemble those observed by JWST, others differ significantly. Notably, the rings in the JWST observations appear almost perfectly circular. Comparable regular features are observed in the simulations when the rings are viewed nearly pole-on near the remnant's center. This alignment suggests that the GM is situated predominantly in the plane of the sky (as also suggested by observations; \citealt{2024ApJ...976L...4D}).

In addition to well-formed, regular ring-hole structures, our simulations also reveal regions pockmarked with overlapping holes, multiple-ring configurations, and partial rings, consistent with JWST observations (\citealt{2024ApJ...976L...4D}). The overlapping holes emerge in areas of high ejecta fragmentation, where "swarms" of ejecta knots puncture the shell at multiple locations, creating a network of interconnected voids and rings (see, for instance, regions 1 and 5 in Fig.~\ref{comparison}). In run W15-IIb-sh-MHD+dec-rl-hr, these features are particularly prominent at the age of \casa, a consequence of the extensive fragmentation driven by radiative losses. However, the frequency of such holes in this phase of evolution appear significantly higher in the simulation compared to the GM inferred from observations.

A scenario more consistent with the GM is found in the simulation at an earlier stage, when the ejecta had only recently begun interacting with the shell ($t = 258$~years; see Fig.~\ref{evol_gm} and regions 1-4 in Fig.~\ref{comparison}). At this stage, the holes are smaller and sparser, aligning more closely with the sizes and distributions observed by JWST in the GM. This comparison suggests that the radius of the shell in our model may need to be increased slightly (by $\approx 25$\%) to delay the interaction between the ejecta and the shell by about 100~years. This adjustment would allow for the formation of holes and rings with morphologies more closely resembling those observed. Another possible explanation is that the shell itself deviates from the idealized spherical shape assumed in our simulations. For instance, the shell could be non-spherical, leading to ejecta interactions occurring at different times across various regions. Alternatively, the shell might consist of multiple layers, with holes forming at different times depending on the specific layer being impacted by the ejecta. This interpretation is supported by evidence of a multi-ring structure detected in position (d3) in Fig.~4 of \cite{2024ApJ...976L...4D}. These complexities in shell geometry and structure could produce features more consistent with the observed GM morphology.

In our simulations, multi-ring structures, such as the one observed in region 4 of Fig.~\ref{comparison}, can also originate from the intricate structure of ejecta clumps, which interact with the shell in a sequential manner. For instance, an initial fragment of a clump might penetrate the shell, forming a primary ring-like structure. Subsequently, a denser fragment from the same clump could impact a slightly offset region, producing additional concentric rings of varying sizes and intensities. These interactions underscore the dynamic interplay between the heterogeneity of the ejecta and the geometry of the shell, illustrating how complex, layered structures can emerge in the circumstellar environment.

Partial rings, on the other hand, form in regions where multiple ejecta knots interact with the shell in close proximity but at slightly staggered times. This temporal delay in interaction can lead to incomplete or perturbed rings and the formation of non-spherical or irregularly shaped holes. These features are particularly evident in areas where ejecta clumps are fragmented or asymmetric, causing the resulting holes and rings to appear distorted or deformed (see Fig.~\ref{comparison}). The resulting diversity in morphologies underscores the sensitivity of the system to variations in both the structure of the ejecta and the local properties of the shell.

Our findings suggest a more intricately structured and inhomogeneous shell, a scenario further supported by the connection between the GM and the QSFs. The QSFs, observed as scattered structures forming an elliptical shape that extends beyond the GM (see Fig.~8 of \citealt{2024ApJ...976L...4D}), likely originate from asymmetric mass loss in the progenitor system. This asymmetry implies that the circumstellar mass is not evenly distributed at a uniform distance from the explosion center, as assumed for the shell in our models. Instead, denser regions of the CSM are located at varying distances, leading to interactions with the remnant at different times. High-angular-resolution observations, such as those collected by JWST, will be essential for refining the geometry and density distribution of the asymmetric shell in our simulations. These constraints will help resolve the discrepancies identified and shed light on the mass-loss events that characterized the final stages of the progenitor's evolution.

Another key consideration is why the GM appears prominently in the JWST observations, whereas the QSFs, despite also representing shocked circumstellar material, are less distinct. One plausible explanation is that the dust, which dominates the infrared emission captured by JWST, may have been destroyed or eroded in other parts of the shell. Dust survival is more likely in regions that have been more recently shocked, further supporting the idea that the remnant's interaction with the GM occurred relatively recently. This timing would preserve the dust in the GM, enhancing its detectability in the infrared. Alternatively, it could also be that the dust in recently shocked regions, such as the GM, is warmer and emits strongly in JWST wavelengths. In contrast, dust in earlier shocked regions, such as the QSFs, may have cooled quickly to lower temperatures, shifting their emission out of the JWST sensitivity range. Additionally, processes such as sputtering and thermal heating over time may have further diminished the visibility of dust in other parts of the shell, reducing their contribution to the observed emission.

   \begin{figure*}
   \centering
   \includegraphics[width=\textwidth]{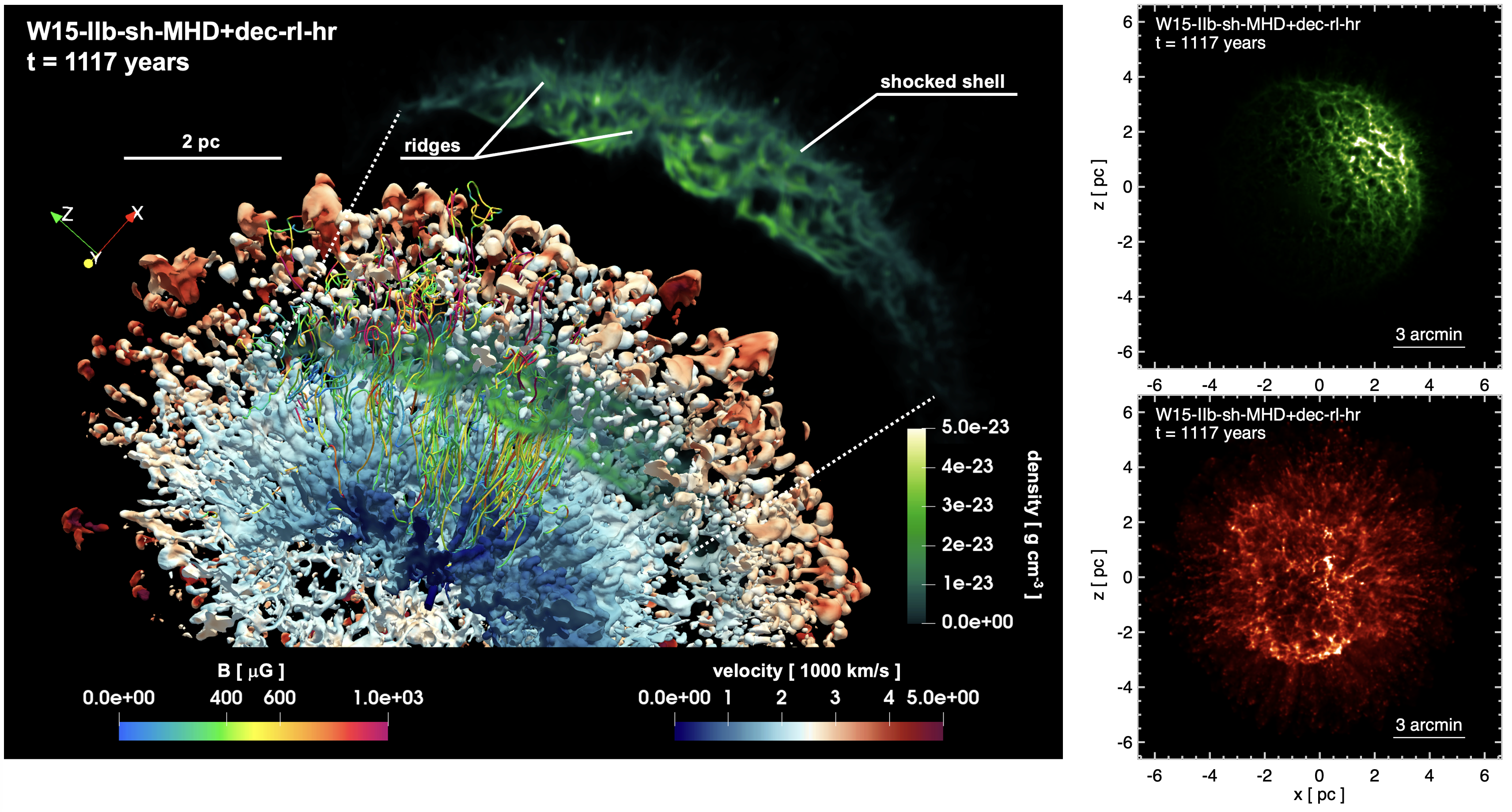}
   \caption{Left panel: Ejecta-shell interaction region from model W15-IIb-sh-MHD+dec-rl-hr at an age of 1117 years, analogous to Fig.~\ref{HD-dec}. The heavily perturbed 3D structure of the shocked shell is also separately displayed as a green volumetric rendering in the upper right corner of the panel. Sampled magnetic field lines are included within a limited region to illustrate the complexity of the magnetic field topology. See online Movie~1 for an animation of these data. A navigable 3D graphic of the remnant structure as well as of the magnetic field configuration in this phase is available at https://skfb.ly/pt7wt. Right panels: Similar to Fig.~\ref{evol_gm}, these panels display the shocked shell (top) and shocked ejecta (bottom) derived from model W15-IIb-sh-MHD+dec-rl-hr at the same evolutionary stage.}
   \label{future_evolution}%
   \end{figure*}

\subsection{Future evolution of the Green Monster}

In the subsequent years, the ejecta continue to expand, intensifying their interaction with the circumstellar shell as ejecta fingers and clumps penetrate deeper into the structure (see the online Movies~1 and 2). This prolonged interaction results in increasingly pronounced perturbations in the shell, particularly evident in model W15-IIb-sh-MHD+dec-rl-hr, where extensive fragmentation of the ejecta, driven by radiative losses, plays a dominant role (see Sect.~\ref{sec:holes_rings}). This evolution gives rise to a progressively intricate pattern of overlapping holes and rings within the shell, along with the formation of pronounced ridges created by the continuous breakthroughs of ejecta knots (see Fig.~\ref{future_evolution} and the online Movies~1 and 2).

During this phase, the mixing between the ejecta and the shell material becomes highly efficient, with turbulent motions facilitating the blending of ejecta material with CSM material. Simultaneously, the magnetic field evolves into a highly complex topology due to perturbations caused by ejecta clumps and fingers (see also \citealt{1996ApJ...472..245J}). This topology is dominated by a prominent radial component, with field strengths reaching up to a few milligauss. The radial structure arises from the combined effects of ejecta fingers stretching the magnetic field lines outward and the inward motion of the reverse shock. This dynamic interaction results in a highly structured and evolving magnetic environment, as illustrated by the sampled magnetic field lines in Fig.~\ref{future_evolution} and in the online Movie~1.

As the interaction progresses, the structure of the shocked shell undergoes substantial changes, diverging significantly from the pockmarked pattern characteristic of the GM. This emphasizes that the GM-like morphology observed in \casa\ is a transient feature, present only during the initial few hundred years of ejecta-shell interaction. Over time, the continued expansion and mixing of the ejecta strongly perturb the distinct holes and rings, transitioning them into a more intricate and complex morphology. 

These findings highlight the critical importance of observing young SNRs like \casa\ to explore the final evolutionary phases of massive progenitor stars. Indeed the transient nature of the GM provides a unique and valuable look into the early stages of ejecta-CSM interaction. Determining the precise timing of this interaction is essential for reconstructing the structure of the CSM at the time of core collapse. This, in turn, can provide critical insights into the progenitor's mass-loss history and the mechanisms responsible for the mass loss in the final stages before the explosion.

\section{Summary and conclusions}
\label{sec:summary}

In this study, we investigated the origin of the GM in the \casa\ SNR and its distinctive structure, as revealed by JWST observations (\citealt{2024ApJ...965L..27M, 2024ApJ...976L...4D}). Using high-resolution 3D MHD simulations, we explored whether the GM could be explained by interactions between the remnant and a dense, asymmetric circumstellar shell, as proposed in previous studies (\citealt{2022A&A...666A...2O}). Additionally, we examined the roles of small-scale ejecta structures, radiative cooling, and magnetic fields in shaping the GM’s morphology, characterized by nearly circular holes and ring-like features. By comparing our simulations with observations from JWST and Chandra, we gained valuable insights into the origins of these features and their broader implications for the final evolutionary stages of \casa’s progenitor star. Our key findings are summarized below.

\begin{itemize}
\item {\bf Origin of Green Monster's pockmarked structure.} 
Our simulations support the interpretation that the GM and its intricate pockmarked structure arise from interactions between ejecta clumps and a dense, asymmetric circumstellar shell. These interactions are driven by HD instabilities, such as Rayleigh-Taylor and Kelvin-Helmholtz instabilities, which generate ejecta fingers and clumps extending from the contact discontinuity. As these structures penetrate the shell, they displace its material, forming nearly spherical holes encircled by dense, ring-like structures.

\item {\bf Role of radiative cooling and magnetic fields.} 
Radiative losses significantly enhance the compression of post-shock ejecta, producing smaller and denser knots and filaments compared to non-radiative models. NEI further amplifies these effects by increasing radiative cooling, which intensifies ejecta fragmentation and produces smaller holes, with sizes as small as $\approx 1^{\prime\prime}$, in good agreement with JWST observations. Furthermore, magnetic fields stabilize smaller-scale structures by partially suppressing HD instabilities at their boundaries and confining plasma, which further amplifies radiative cooling effects. Together, the combination of efficient radiative cooling and magnetic confinement successfully reproduces the intricate pockmarked patterns observed in \casa, providing a more realistic description of its morphology. 

\item {\bf Comparison with Observations.} 
The simulations successfully reproduce several morphological characteristics of the GM, including overlapping holes, multiple-ring configurations, and partial rings. The inclusion of radiative losses (e.g., in run W15-IIb-sh-MHD+dec-rl-hr) yields structures consistent with JWST observations, though the simulation also predicts some larger holes, with sizes up to $\approx 12^{\prime\prime}$, exceeding those observed. A better match to the GM morphology is achieved at earlier epochs (e.g., 258 years post-explosion), indicating that the timing and configuration of ejecta-shell interactions are critical in shaping these features. This could be refined by considering a dense shell in the pre-SN CSM located at a distance of 1.9~pc from the center of explosion, rather than the 1.5~pc used in our simulations. According to this interpretation, the remnant encountered the CSM layer responsible for the GM approximately $50-100$~years ago. This timing is consistent with the ionization age of $\approx 1.5 \times 10^{11}$~cm$^{-3}$~s in the GM inferred from X-ray observations (\citealt{2024ApJ...964L..11V}), given that the shocked shell in our model exhibits densities ranging between 30 and 150~cm$^{-3}$. Hence, a more realistic circumstellar shell geometry, deviating from a simple spherical shape or consisting of multiple layers, may further improve consistency with the observed GM morphology.

\item {\bf Additional Diagnostics for Future Observations.} The simulations highlight several distinctive properties of the holes and rings produced by the described mechanism. The holes are filled with fingers of ejecta, suggesting that observations should detect a composition contrast between the material inside and around the holes. In particular, we suggest that the gas filling the holes is enriched of He, C, O, and Ne. In the densest portion of the shell, the mass density of the shocked ejecta filling the holes is, on average, about an order of magnitude smaller than that of the shocked shell (compare the top and center panels in Fig.~\ref{holes_struct}). Additionally, the shocked ejecta filling the holes exhibit LoS velocities of approximately $-4000$~km~s$^{-1}$, compared to the shell's LoS velocity of around $-2000$~km~s$^{-1}$ (see Fig.~\ref{distrib_velocity}). Finally, the displaced shell material at the edges of the holes is expected to have tangential velocities (i.e., velocities perpendicular to the LoS) ranging from 200 to 800~km~s$^{-1}$ (see Fig.~\ref{tangent_vel}). These distinctive characteristics of the holes and ring-like structures can be valuable observational markers that can help confirm the mechanism responsible for their formation and also distinguish this process from the alternative scenario involving small, high-velocity ejecta bullets impacting the GM prior to the arrival of the forward shock (see \citealt{2024ApJ...976L...4D}).

\item {\bf Future evolution.} 
As \casa\ evolves, simulations predict that ejecta fingers and clumps will penetrate deeper into the circumstellar shell, heavily perturbing the shocked shell and forming an intricate network of overlapping holes, rings, and the formation of ridges. Turbulent mixing will continue to blend ejecta with circumstellar material efficiently. Over time, the distinct GM-like morphology will transition into a more complex and irregular structure, underscoring its transient nature. This may facilitate determining the exact timing of the ejecta-shell interaction for reconstructing the mass-loss history of the progenitor star and understanding the mechanisms that shaped its circumstellar environment.
\end{itemize}

In light of our findings, the GM and its intricate structure serve as a unique diagnostic tool for probing the late evolutionary stages of \casa’s progenitor star. It offers a rare opportunity to investigate the mechanisms that led to the stripping of the progenitor’s H envelope and, more broadly, to explore the poorly understood processes driving mass loss during the final stages of massive star evolution. Specifically, analyzing the geometry of the circumstellar shell (its thickness, spatial distribution, and distance from the progenitor) can provide critical constraints on the timing and nature of the mass-loss episodes. These insights can help identify the underlying mechanism, whether it be eruptive mass loss, binary interactions, or other processes, shedding light on the physical phenomena that shaped the circumstellar environment before the explosion.

If the origin of the circumstellar shell is "intrinsic" to the progenitor star, resulting from the progenitor star’s late-stage mass loss, our model suggests it formed through a violent eruption approximately $10,000-100,000$ years before the SN explosion (\citealt{2022A&A...666A...2O}). This scenario implies an ejection velocity of either a few hundred km~s$^{-1}$ (potentially during a common-envelope phase in a binary system) or $10-20$~km~s$^{-1}$ (characteristic of a red supergiant wind). These estimates are consistent with observations indicating interactions with dense CSM (e.g., \citealt{2018ApJ...866..139K, 2020NatAs...4..584K, 2024ApJ...976L...4D}). Interestingly, binary interaction could have played a role in shaping the circumstellar environment, as proposed in cases where mass transfer or common-envelope evolution leads to significant pre-SN outflows (e.g., \citealt{2012ApJ...752L...2C, 2013A&ARv..21...59I, 2017MNRAS.471.4839S}). Furthermore, our findings align with \citet{2009A&A...503..495V}, who concluded that \casa’s progenitor did not undergo a Wolf-Rayet (WR) phase, as the QSF in \casa\ are inconsistent with shocked WR shell clumps. Since a brief WR phase ($\leq 15,000$ yr) would have left detectable clumps within the SNR, the structured and asymmetric CSM in our model points to an alternative mass-loss history dominated by episodic outflows rather than a steady WR wind. Our results further indicate that the CSM’s complexity exceeds current model representations, suggesting a more intricate and heterogeneous pre-SN environment.

Alternatively, the shell may have an "external" origin. Shell asymmetries, such as those described in our models, could also result from interactions between stellar winds and the interstellar medium (ISM) in runaway massive stars. These stars, moving supersonically through the ISM, generate lopsided bow shock nebulae as a result of wind-ISM interactions (e.g., \citealt{2017MNRAS.464.3229M, 2020MNRAS.496.3906M, 2021MNRAS.506.5170M}). Notably, runaway stars often originate from the breakup of binary systems following the SN explosion of one companion (e.g., \citealt{1961BAN....15..265B, 2001A&A...365...49H, 2015MNRAS.448.3196D}) or from dynamical interactions in dense stellar environments (e.g., \citealt{1986ApJS...61..419G, 2003ARA&A..41...57L, 2011MNRAS.410..304G}). Therefore, determining whether \casa\ is interacting with an asymmetric circumstellar shell reminiscent of those observed around runaway massive stars could provide valuable insights into the mechanisms that stripped the \casa\ progenitor of its H envelope and the absence of a detectable companion star.

However, the irregular structure of the GM and the aspherical shell required by our simulations suggest that a lopsided bow shock nebula, typical of a runaway star, is unlikely. Moreover, the QSFs appear to be closely related to the GM and exhibit chemical signatures of CNO-processed material (e.g., \citealt{1995ApJ...440..706R, 2018ApJ...866..139K}). These signatures strongly suggest an origin tied to the progenitor's mass loss rather than an external ISM interaction. Explaining the presence of CNO-processed material in an "external origin" scenario would be highly challenging, further supporting the interpretation that the shell and associated structures are intrinsic to the progenitor system.

Finally, our simulations highlight the critical role of spatial resolution in comparing models with high-angular-resolution observations, such as those collected with JWST. However, we also find that even our simulations (despite their very high resolution) are at the limit of resolving the smallest observed holes, which have sizes on the order of $1^{\prime\prime}$ (corresponding to $\approx 0.015$~pc at the distance of \casa). This limitation suggests the need for further refinement. Future work, therefore, should focus on increasing spatial resolution to better capture small-scale structures and dynamics. Additionally, exploring the effects of non-spherical shell geometries into the simulations, guided by constraints from high-angular-resolution observations, will be essential to addressing remaining discrepancies. These advancements will refine the models and significantly enhance our ability to interpret observations of \casa\ and, more in general, similar SNRs, providing deeper insights into the physics of ejecta-CSM interactions and the progenitors of core-collapse SNe.

\begin{acknowledgements}
We thank an anonymous referee for the useful suggestions that allowed us to improve the manuscript. S.O. expresses sincere gratitude to Massimiliano Guarrasi and the CINECA team for their invaluable support in using the high-performance computing (HPC) facilities at CINECA. The high-resolution simulations presented in this work would not have been possible without their expertise and the availability of substantial computational resources. Specifically, run W15-IIb-sh-HD+dec-hr required approximately 8 million CPU hours, while run W15-IIb-sh-MHD+dec-rl-hr consumed around 20 million CPU hours on the Leonardo supercomputing facility. The \PLUTO\ code is developed at the Turin Astronomical Observatory (Italy) in collaboration with the Department of General Physics of  Turin University (Italy) and the SCAI Department of CINECA (Italy). We acknowledge the CINECA ISCRA Award N.HP10BUMIQR for the availability of HPC resources and support at the infrastructure Leonardo based in Italy at CINECA. Additional computations were carried out on the HPC system MEUSA at the SCAN (Sistema di Calcolo per l'Astrofisica Numerica) facility for HPC at INAF-Osservatorio Astronomico di Palermo. Computer resources for this project have also been provided by the Max Planck Computing and Data Facility (MPCDF) on the HPC systems Cobra and Draco. The navigable 3D graphics have been developed in the framework of the project 3DMAP-VR (3-Dimensional Modeling of Astrophysical Phenomena in Virtual Reality; \citealt{2019RNAAS...3..176O, 2023MmSAI..94a..13O}) at INAF-Osservatorio Astronomico di Palermo. 
S.O., M.M., F.B., V.S., and E.G. acknowledge financial contribution from the PRIN 2022 (20224MNC5A) - ``Life, death and after-death of massive stars'' funded by European Union – Next Generation EU, and the INAF Theory Grant ``Supernova remnants as probes for the structure and mass-loss history of the progenitor systems''. 
H.-T.J.\ acknowledges support by the German Research Foundation (DFG) through the Collaborative Research Centre ``Neutrinos and Dark Matter in Astro- and Particle Physics (NDM),'' grant No. SFB-1258-283604770, and under Germany's Excellence Strategy through the Cluster of Excellence ORIGINS EXC-2094-390783311. 
D.M. acknowledges support from the National Science Foundation through grants PHY-2209451 and AST-2206532. 
I.D.L. acknowledges funding from the Belgian Science Policy Office (BELSPO) through the PRODEX project ``JWST/MIRI Science exploitation'' (C4000142239) and funding from the European Research Council (ERC) under the European Union's Horizon 2020 research and innovation program DustOrigin (ERC-2019-StG-851622).
S.N. acknowledges support by the ASPIRE project for top scientists, JST RIKEN-Berkeley Mathematical Quantum Science Initiative, and from ``Inter- disciplinary Theoretical and Mathematical Sciences Program of RIKEN (report number: RIKEN-iTHEMS-Report-25).''
T.T. acknowledges support from the NSF grant AST-2205314 and the NASA ADAP award 80NSSC23K1130.
\end{acknowledgements}

\bibliographystyle{aa}
\bibliography{references}

\clearpage
\begin{appendix}
\onecolumn

\section{The radiative losses from optically thin plasma}
\label{app:losses}

In our simulations, radiative losses are calculated for each cell in the simulation domain using a table lookup/interpolation scheme. This scheme provides the radiative cooling term, $\Lambda(T_{\rm e}, \tau, Z)$, as a function of the electron temperature ($T_{\rm e}$), ionization age ($\tau$), and abundance ($Z$) of the emitting plasma. The lookup table was computed using SPEX version 3.08.01 (\citealt{1996uxsa.conf..411K, 2018zndo...2419563K}), including the latest updates to the SPEX Atomic Code and Tables (SPEXACT v3.08.01). These updates include the most recent radiative loss rates and cooling rates available in the SPEXACT database and associated routines. More specifically, the radiative losses in the table were calculated over a grid with 100 logarithmically spaced temperature bins, covering the range 0.001 to 100 keV, and 100 logarithmically spaced ionization time bins, spanning $10^8$ to $10^{13}$~cm$^{-3}$~s. 

\begin{figure*}[!hb]
   \centering
   \includegraphics[width=0.98\textwidth]{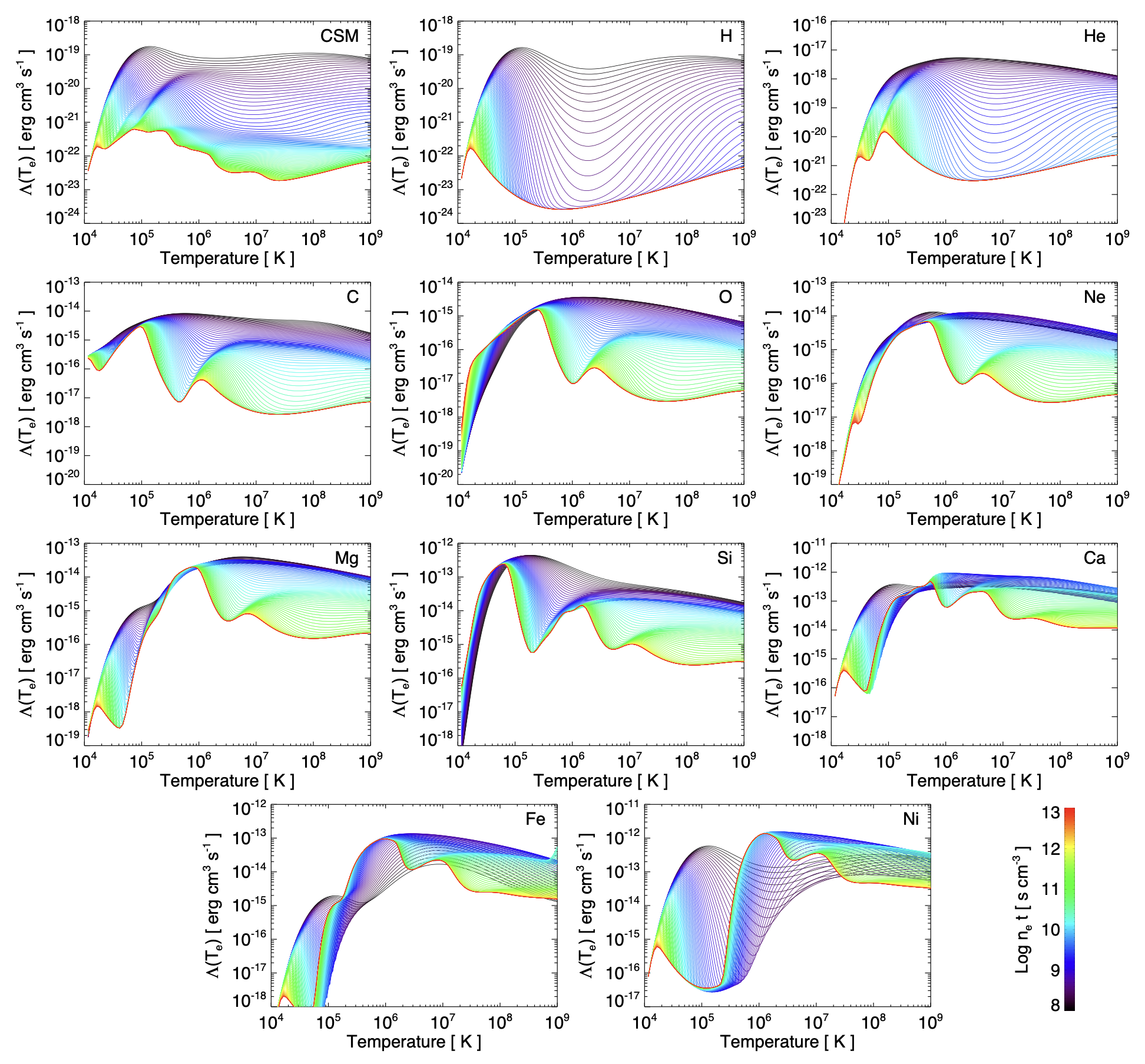}
   \caption{Radiative losses from optically thin plasma as a function of electron temperature for selected values of ionization age, derived from SPEX version 3.08.01 (\citealt{1996uxsa.conf..411K, 2018zndo...2419563K}). The upper left panel illustrates the cooling function assuming solar metal abundances from \cite{2009LanB...4B..712L}, highlighting the dependence on ionization time. The remaining panels show radiative losses for plasmas composed of the individual elements indicated in the labels.}
   \label{rad_losses}%
   \end{figure*}

Figure~\ref{rad_losses} illustrates $\Lambda(T_{\rm e}, \tau, Z)$ as a function of $T_{\rm e}$ for selected values of $\tau$. Red curves in the figure corresponds to plasmas in CIE. The upper left panel shows radiative losses for a plasma with solar metal abundances from \cite{2009LanB...4B..712L}, following the methodology described in \cite{2021A&A...655A...2S}. The remaining panels display radiative losses for plasmas composed of single elements, providing insight into how individual species contribute to the total cooling. These last curves were used to reconstruct the radiative cooling term in the ejecta. As a validation step, we confirmed that our cooling functions can closely reproduce those of \cite{2021A&A...655A...2S} across the range 0.01 to 100 keV for all elements relevant to this work.

In the table lookup/interpolation scheme, the values of $T_{\rm e}$ and $\tau$ required to compute radiative losses in a given cell of the domain are determined as described in previous works (\citealt{2015ApJ...810..168O, 2016ApJ...822...22O}). The electron temperature is calculated in each cell by assuming rapid electron heating at shock fronts up to $kT = 0.3$~keV (\citealt{2007ApJ...654L..69G}) and applying the effects of Coulomb collisions to determine ion and electron temperatures in the post-shock plasma of the cell (see \citealt{2015ApJ...810..168O} for more details). The values of $\tau$ are calculated in each cell as $\tau = n_{\rm e} \delta t$, where $n_{\rm e}$ is the electron density in the cell, and $\delta t$ is the time elapsed since the plasma in the cell was shocked. This approach ensures computational efficiency (especially for 3D simulations like ours) while maintaining reasonable accuracy in evaluating NEI effects (e.g., \citealt{2010MNRAS.407..812D}).

For the shocked CSM, radiative losses are computed assuming solar metal abundances (\citealt{2009LanB...4B..712L}), as shown in the upper right panel of Fig.~\ref{rad_losses}. For the shocked ejecta, radiative losses depend on the chemical composition of the plasma in each cell. This composition is tracked using passive tracers that follow the evolution of elements synthesized during the SN explosion. In our case, the adopted SN simulation (model W15-2-cw-IIb; \citealt{2017ApJ...842...13W}) incorporates a nuclear reaction network that includes 11 species: protons (1H), $^4$He, $^{12}$C, $^{16}$O, $^{20}$Ne, $^{24}$Mg, $^{28}$Si, $^{40}$Ca, $^{44}$Ti, $^{56}$Ni, and an additional “tracer nucleus” $^{56}$X, representing Fe-group species synthesized in neutron-rich environments, such as neutrino-heated ejecta.

The radiative cooling term for the shocked ejecta, therefore, is computed in each cell from the 11 species considered as:

\begin{equation}
{\rm cooling\, term} = \sum_{\rm el} n_{\rm e,el} n_{\rm i,el} \Lambda_{\rm el} (T_{\rm e}, \tau)
\end{equation}

\noindent
where $n_{\rm e,el}$ and $n_{\rm i,el}$ are the number densities of free electrons and ions, respectively, for element "el" in the cell and $\Lambda_{\rm el} (T_{\rm e}, \tau)$ is the radiative cooling function for element "el" as a function of $T_{\rm e}$ and $\tau$, shown in Fig.~\ref{rad_losses}.

The radiative losses are incorporated into the numerical code \PLUTO\ using a fractional step formalism (\citealt{2007ApJS..170..228M}), wherein the HD evolution and the radiative source term are treated separately through operator splitting. This approach ensures numerical stability and accuracy, as the source and advection steps are decoupled. This method preserves second-order accuracy in time, provided that both the advection step and the source step are independently at least second-order accurate (see \citealt{2007ApJS..170..228M} for more detail). The use of operator splitting also allows for efficient computation of radiative cooling, even in cases where the cooling timescales vary significantly across the domain, as it is the case for the simulations performed in this paper.

\begin{table}
\caption{Setup for the additional test simulations.}
\label{Tab:test}
\begin{center}
\begin{tabular}{llll}
\hline
\hline
SNR Model                    & {\bf B} & radiative & grid \\ 
                             &         & losses      \\ \hline
W15-IIb-sh-MHD+dec-rl        & yes  & NEI      & $(1024)^3$ \\
W15-IIb-sh-HD+dec-CIE-hr     & no   & CIE      & $(2048)^3$  \\
W15-IIb-sh-MHD+dec-CIE-hr    & yes  & CIE      & $(2048)^3$ \\
\hline
\end{tabular}
\end{center}
\end{table}

Additional test simulations were performed to evaluate the role of spatial resolution and the influence of NEI on the radiative cooling process. The properties of these test simulations are summarized in Table~\ref{Tab:test}.

\begin{figure*}
   \centering
   \includegraphics[width=0.98\textwidth]{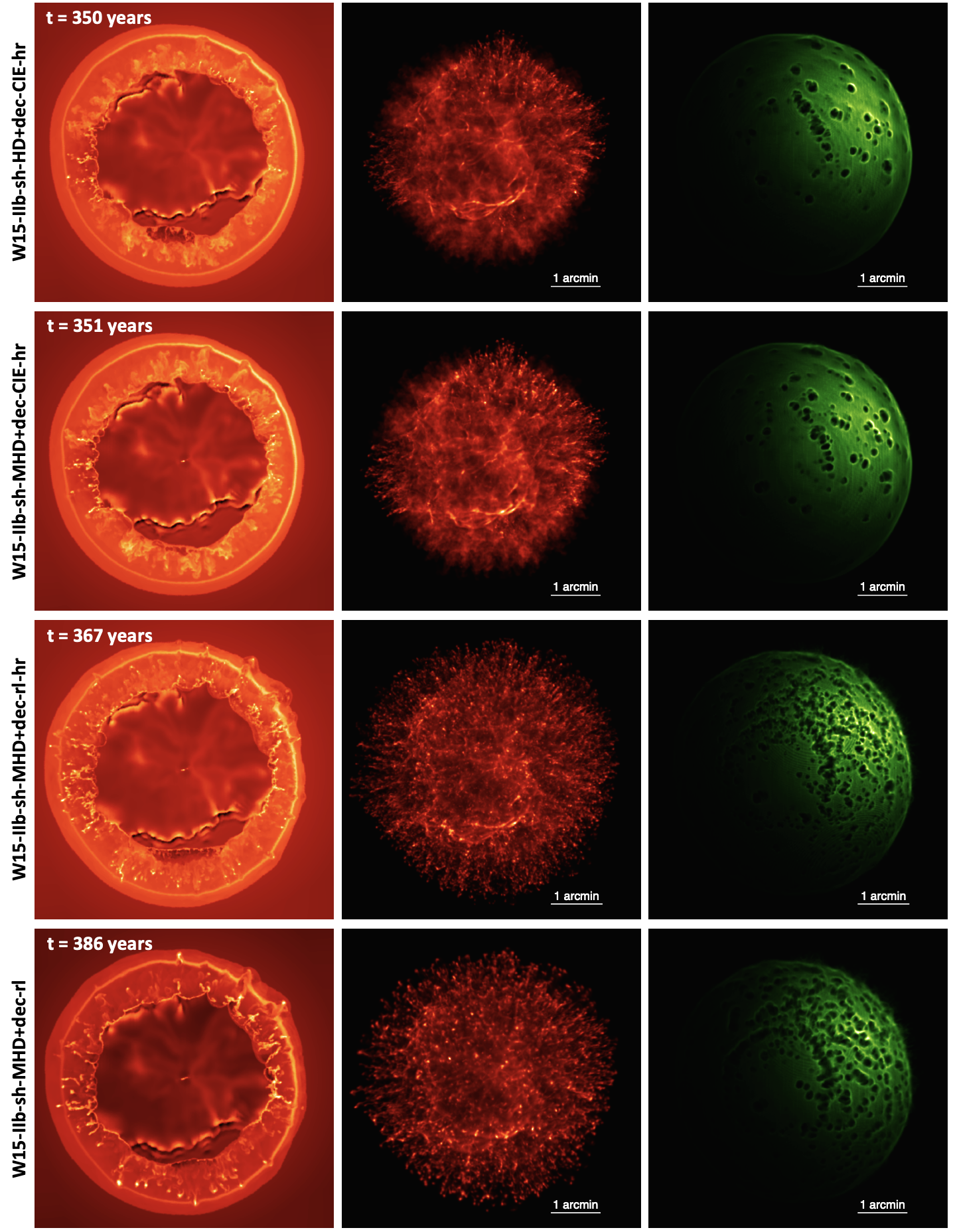}
   \caption{Comparison of density distributions from the test simulations reported in Table~\ref{Tab:test} and model W15-IIb-sh-MHD+dec-rl-hr (Table~\ref{Tab:model}). The figure presents 2D slices in the $(x, z)$ plane showing the plasma density distribution in logarithmic scale (left panels), along with 3D volumetric renderings of the ejecta (center panels) and shell (right panels) at the labeled times (see upper left corner in the left panels).}
   \label{app_test}%
   \end{figure*}

To evaluate the effect of spatial resolution on cooling and the formation of dense ejecta fingers and clumps, we compared the high-resolution simulation run W15-IIb-sh-MHD+dec-rl-hr (Table~\ref{Tab:model}) with a lower-resolution counterpart performed on a $1024^3$ grid, using the same physical setup. The results of this comparison are shown in the last two rows in Fig.~\ref{app_test}. In both cases, dense, thin ejecta fingers form and fragment into clumps, but the lower-resolution simulation produces fewer and larger clumps compared to the high-resolution run (see the 2D slices of density distribution shown in the left panels of the figure for these two simulations). Smaller clumps, which are well-resolved in the high-resolution simulation, are smoothed out or absent in the lower-resolution model. This difference is due to the superior ability of the high-resolution simulation to capture the collapse of ejecta under radiative cooling, resolving fine structures that are lost at lower resolution. Since this study aims to replicate features as small as $0.015$~pc observed with JWST, achieving high spatial resolution is critical for accurately modeling radiative cooling and for a meaningful comparison with observations.

Figure~\ref{app_test} also illustrates the impact of the magnetic field on enhancing the radiative cooling of the ejecta. The first and second rows compare models that differ only in the presence of a magnetic field, with run W15-IIb-sh-MHD+dec-CIE-hr (second row) including magnetic effects. The density distributions in the two simulations are broadly similar, but in W15-IIb-sh-MHD+dec-CIE-hr, the Fe-rich ejecta clumps undergoing strong radiative cooling are denser and more compact compared to those in W15-IIb-sh-HD+dec-CIE-hr (first row). This difference is evident in the shocked Fe-rich plume expanding to the left and interacting with the reverse shock. The differences can be clearly observed in the 2D slices of density distribution shown in the left panels for the two simulations. These findings highlight the role of magnetic fields in confining ejecta clumps, thereby increasing their densities and enhancing cooling efficiency.

Finally, we evaluated the impact of NEI effects on the simulations by comparing run W15-IIb-sh-MHD+dec-rl-hr with a simulation that considers the same setup but where radiative losses were calculated under the assumption of CIE (run W15-IIb-sh-MHD+dec-CIE-hr; see Table~\ref{Tab:test}). The results indicate that assuming CIE significantly underestimates radiative losses (red curves in Fig.~\ref{rad_losses}), especially for light and intermediate-mass elements (H, He, C, O, Ne, Mg, and Si), which in turn affects the simulation outcomes. Figure~\ref{app_test} highlights the differences in density distributions between simulations either with or without NEI effects. In the CIE-based run W15-IIb-sh-MHD+dec-CIE-hr (second row), the level of ejecta fragmentation and the number of isolated clumps are notably lower compared to the NEI-inclusive run W15-IIb-sh-MHD+dec-rl-hr (third row). These discrepancies arise because, in the CIE-based simulation, plasma with electron temperatures in the range of tens of millions of Kelvin is incorrectly assumed to be in equilibrium. This assumption severely underestimates the contribution of emission lines to radiative losses, particularly for intermediate-mass elements such as C, O, Ne, Mg, and Si. In fact, the distributions of these intermediate-mass elements show negligible differences between the CIE-based model and the model without radiative losses (e.g., run W15-IIb-sh-HD+dec-hr in Table~\ref{Tab:model}; compare Fig.~\ref{app_test} and Fig.~\ref{evol_gm}). This demonstrates that neglecting NEI effects compromises the accuracy of modeling radiative cooling in freshly shocked plasma, particularly for capturing the fragmentation, clumping, and mixing processes that shape the physical and observational properties of \casa.

\section{Online multi-media material}
\label{app:multi-media}

This paper is accompanied by a collection of online movies and interactive 3D graphics that provide detailed visualizations of the shocked shell structure and its interaction with the remnant. These supplementary materials aim to enhance the reader's understanding by offering an interactive and dynamic exploration of the shell's evolution over time. Below, we present a detailed description of these resources.

\medskip
\noindent
{\bf Online Movies.} The following movies are available as supplementary material on the A\&A webpage:

\begin{itemize}

\item Movie 1: "Remnant-shell interaction". This movie complements Fig.~\ref{MHD-dec-rl}, illustrating the interaction between the SNR and the circumstellar shell. It spans the evolution from the moment the forward shock impacts the shell at $t\sim 200$~years to the remnant stage at $t\sim 1100$~years. The upper-right inset provides a detailed view of the shell structure, showcasing the formation of holes and rings over time. To enhance clarity, the magnetic field lines are displayed in only a portion of the volume, showcasing the complexity of the magnetic field configuration while allowing for an clearer inspection of the ejecta structure in other regions of the remnant.

\item Movie 2: "Production of Holes and Rings in Region 6". This movie complements Fig.~\ref{holes_struct}, providing a detailed close-up view of the interaction between ejecta clumps and fingers and the circumstellar shell in region 6 (see also Fig.~\ref{comparison}) in the period between $t\sim 200$~years and $t\sim 1100$~years. The panels on the right offer separate views of the shocked shell, observed from two different lines-of-sight, enabling a clearer perspective on its evolution, structural disruption, and the emergence of distinctive features. The movie highlights how the ejecta penetrate and distort the shell, revealing the intricate interplay between the ejecta dynamics and the shell's response over time.

\end{itemize}

\noindent
{\bf Interactive 3D Graphics.} These models are available as supplementary material on the A\&A webpage. Additionally they are hosted on the Sketchfab platform and are compatible with virtual reality devices (3DMAP-VR project; \citealt{2019RNAAS...3..176O, 2023MmSAI..94a..13O}). They provide an interactive experience to complement the figures in the paper, allowing users to zoom, pan, and rotate views. Each model includes labeled annotations to highlight key features:

\begin{itemize}
\item Model 1: "Origin of the Green Monster in Cassiopeia A". Accessible at \url{https://skfb.ly/psYpK}, this model complements Fig.~\ref{MHD-dec-rl}. It allows users to explore the complex structure of the ejecta, the shocked shell, and the perturbed magnetic field at the age of \casa. Key features include: (1) Ejecta structure; (2) Perturbed magnetic field; (3) Shocked shell; (4) Separate display of the shocked shell.

\item Model 2: "Formation of Holes and Rings in Cassiopeia A". Accessible at \url{https://skfb.ly/ptFHY}, this model complements Fig.~\ref{holes_struct}. It illustrates the formation of holes and rings during the interaction between the ejecta and the shocked shell. Key features include: (1) Circumstellar shell; (2) Ejecta material fingers; (3) Magnetic field; (4) Ejecta clumps; (5) Shell disruption.

\item Model 3: "A millennium-old remnant of a neutrino-driven SN".  Accessible at \url{https://skfb.ly/pt7wt}, this model complements Fig.~\ref{future_evolution} and visualizes the ejecta and shocked shell structure approximately 1000 years after the SN event. Key features include: (1) Ejecta structure; (2) Magnetic field configuration; (3) Shocked shell structure.
\end{itemize}

These multimedia resources are integral to the study, offering a more comprehensive and immersive perspective on the phenomena discussed. We encourage readers to explore these materials for a deeper understanding of the results presented in the paper.

\end{appendix}

\end{document}